\documentclass[%
preprint,
 amsmath,amssymb,
 aps,
prb,
floatfix,
]{revtex4-1}

\usepackage{graphicx}
\usepackage{dcolumn}
\usepackage{bm}

\begin{document}


\title{Thermal Transport Across Metal Silicide-Silicon Interfaces: An Experimental Comparison between Epitaxial and Non-epitaxial Interfaces}

\author{Ning Ye}
\affiliation{%
 Department of Mechanical Engineering, University of Delaware 130 Academy St., Newark, DE, 19716, USA
}%
\author{Sridhar Sadasivam}%
\affiliation{%
 School of Mechanical Engineering and Birck Nanotechnology Center, Purdue University, 1205 W. State St., West Lafayette, IN  47907 USA
}%
\author{Timothy S. Fisher}%
\affiliation{%
 School of Mechanical Engineering and Birck Nanotechnology Center, Purdue University, 1205 W. State St., West Lafayette, IN  47907 USA
}%
\author{Joseph P Feser}
 \email{jpfeser@udel.edu}
\affiliation{%
 Department of Mechanical Engineering, University of Delaware 130 Academy St., Newark, DE, 19716, USA
}%
\author{Tianshi Wang}
\affiliation{%
 Materials Science and Engineering Department, University of Delaware, 201 Dupont Hall, Newark, DE, 19716, USA
}%
\author{Chaoying Ni}
\affiliation{%
 Materials Science and Engineering Department, University of Delaware, 201 Dupont Hall, Newark, DE, 19716, USA
}%
\author{Anderson Janotti}
\affiliation{%
 Materials Science and Engineering Department, University of Delaware, 201 Dupont Hall, Newark, DE, 19716, USA
}%

\date{\today}

\begin{abstract}
Silicides are used extensively in nano- and microdevices due to their low electrical resistivity, low contact resistance to silicon, and their process compatibility. In this work, the thermal interface conductance of TiSi$_2$, CoSi$_2$, NiSi and PtSi are studied using time-domain thermoreflectance. Exploiting the fact that most silicides formed on Si(111) substrates grow epitaxially, while most silicides on Si(100) do not, we study the effect of epitaxy, and show that for a wide variety of interfaces there is no dependence of interface conductance on the detailed structure of the interface. In particular, there is no difference in the thermal interface conductance between epitaxial and non-epitaxial silicide/silicon interfaces, nor between epitaxial interfaces with different interface orientations. While these silicide-based interfaces yield the highest reported interface conductances of any known interface with silicon, none of the interfaces studied are found to operate close to the phonon radiation limit, indicating that phonon transmission coefficients are non-unity in all cases and yet remain insensitive to interfacial structure.  In the case of CoSi$_2$, a comparison is made with detailed computational models using (1) full-dispersion diffuse mismatch modeling (DMM) including the effect of near-interfacial strain, and (2) an atomistic Green' function (AGF) approach that integrates near-interface changes in the interatomic force constants obtained through density functional perturbation theory.  Above 100K the AGF approach significantly underpredicts interface conductance suggesting that energy transport does not occur purely by coherent transmission of phonons, even for epitaxial interfaces.  The full-dispersion DMM closely predicts the experimentally observed interface conductances for CoSi$_2$, NiSi, and TiSi$_2$ interfaces, while it remains an open question whether inelastic scattering, cross-interfacial electron-phonon coupling, or other mechanisms could also account for the high temperature behavior.  The effect of degenerate semiconductor dopant concentration on metal-semiconductor thermal interface conductance was also investigated with the result that we have found no dependencies of the thermal interface conductances up to (n-type or p-type) $\approx 1\times10^{19}$ cm$^{-3}$, indicating that there is no significant direct electronic transport and no transport effects which depend on long-range metal-semiconductor band alignment.
\end{abstract}

\maketitle
\section{\label{sec:level1}Introduction}
Metal silicide thin films are present in nearly all modern silicon microelectronic devices.  In particular, the silicides PtSi, WSi$_2$, TiSi$_2$, CoSi$_2$, NiSi are used extensively due to their low electrical contact resistance to Si, low resistivity, and chemical process compatibility, as well as the low thermal budget associated with their formation.\cite{RN1,RN2}  They can serve a wide range of roles including ohmic contacts, Schottky barrier contacts, gate electrodes, local interconnects, and diffusion barriers. While many silicides are excellent thermal conductors due to their low electronic resistivity, they are generally applied as thin films with nanoscale thicknesses, such that interfacial properties are expected to dominate thermal transport locally.\cite{RN3,RN4,RN5}  This work reports the experimental measurements of thermal interface conductance on a wide range of technologically relevant metallic silicide-silicon interfaces, and shows that they are the highest thermal interface conductances ever measured for a metal-silicon interface on silicon, and are comparable to the highest metal-dielectric thermal interface conductances ever measured.

In addition to the practical implications to thermal management in microelectronics, silicide interfaces represent a unique opportunity for studying the fundamental physics of thermal transport across interfaces.  In general, a lack of experimental data exists regarding the role of disorder on thermal interface conductance, and in particular data for which the interfacial structure is known is scarce.  Despite a large number of investigations of thermal interface conductance in literature, there are just a few which directly measure the thermal conductance of epitaxial metals on crystalline substrates.  Stoner and Maris have reported the thermal interface conductance of single crystal Au grown perpendicular to [2110] Al$_2$O$_3$, and found it to be more than 3 times higher than for similar non-epitaxial samples.  Compared to theory, epitaxial Au/Al$_2$O$_3$ interfacial conductance greatly exceeded lattice dynamics calculations and some measurements were in excess of the phonon radiation limit\cite{PhysRevB.48.16373}.  Costescu et al \cite{RN6} have measured TiN thin films grown on MgO and Al$_2$O$_3$ substrates.  They compared thermal interface conductance of epitaxial TiN/MgO(001) to TiN/MgO(111) growth and found no difference in their values, in spite of the large number of stacking faults in the latter case.  Their data neither fit a coherent lattice dynamics model (which overestimated conductance by $\approx$70\%), nor a Debye-based diffuse mismatch model (which overestimated by $\approx$300\%), though if optical modes were excluded from mode-conversion better agreement was found ($\approx$50\%).  Wilson\cite{Wilson_PhysRevB.91.115414} has studied the thermal interface conductance of epitaxial SrRuO$_3$ grown on SrTiO$_3$ and estimated a lower bound of  $G_{\textrm{SrRuO$_3$/SrTiO$_3$}}\approx$800 MW/m$^2$-K.  In the case of interfaces with silicon, only a few reports of direct measurements of interface conductance exist in which the substrate was cleaned and the oxide was removed prior to interface formation; thus, the detailed lattice structure of the interface is typically unclear, and the presence of native oxide is virtually assured.  A couple reports\cite{Minnich_PhysRevLett.107.095901,RN24} of thermal interface conductance measurements of polycrystalline Al(111) growth on HF dipped Si(100) substrates exist and show that clean interfaces have substantially higher thermal interface conductance than untreated surfaces.  However, epitaxial Al does not readily form on Si, and thus comparisons between experiment and theories considering interface structure are still lacking. Liu \cite{RN7} recently reported the first thermal conductance measurement of an epitaxial metal with silicon:  a NiSi$_2$/Si interface within a Si nanowire created by a reactive method using an in-situ electron beam heating technique. The thermal interface conductance reported was unusually high:  G = 500 MW/m$^2$-K at 300K.   Taken together, these measurements show that epitaxial interfaces can produce record-breaking phonon-dominated thermal interface conductances.  References \onlinecite{PhysRevB.48.16373,RN6,RN7,Wilson_PhysRevB.91.115414} also serve as the only metal-dielectric interface thermal conductance measurements performed where the interfaces were simultaneously structurally and thermally characterized. Thus, to date they are the only experiments with which direct theoretical comparisons can be made.  Despite this, it would appear that no such comparisons have been made using modern computational tools.  Consequently, there are substantial open questions about the physics of transport across epitaxial as well as disordered interfaces.  

Silicide-silicon interfaces are a great testing platform for the effect of disorder on thermal interface conductance because many metal silicide interfaces can be grown either as epitaxial or non-epitaxial interfaces depending on the synthetic process and substrate orientation. Also, many silicides are metallic and optically opaque, which enables their direct use in modern optical thermal interface conductance characterization methods such as time-domain thermoreflectance (TDTR). Epitaxial growth of metal silicides on silicon has been previously demonstrated for most known silicides including PtSi\cite{Ishiwara_JAP_1979_PtSi}, CoSi$_2$\cite{Dass_APL_1991,DAvitaya_JVac_1985}, NiSi\cite{Mangelinck_APL_1999_75_12}, C54 TiSi$_2$\cite{WAN1997105}, C49 TiSi$_2$\cite{Yang_EpiC49_JAP_2003}, VSi$_2$\cite{:/content/aip/journal/jap/57/6/10.1063/1.334420}, CrSi$_2$\cite{Kim19998}, $\gamma$- \& $\beta$-FeSi$_2$\cite{GRIMALDI199419},  YSi$_2$\cite{Gurvitch_APL_1987}, YSi$_2$\cite{Knapp_APL_1986_REsilicides}, GdSi$_2$\cite{Knapp_APL_1986_REsilicides}, TbSi$_2$\cite{Knapp_APL_1986_REsilicides}, DySi$_2$\cite{Knapp_APL_1986_REsilicides}, HoSi$_2$\cite{Knapp_APL_1986_REsilicides}, ErSi$_2$\cite{Knapp_APL_1986_REsilicides}, TmSi$_2$\cite{Knapp_APL_1986_REsilicides}, YbSi$_2$\cite{Knapp_APL_1986_REsilicides}, LuSi$_2$\cite{Knapp_APL_1986_REsilicides}, MoSi$_2$\cite{Lin_APL_epiMoSi2}, Pd$_2$Si\cite{KIRCHER1971507}, TaSi$_2$\cite{Wu_JAP_1987_TaSi2}, WSi$_2$\cite{Lin_JAP_1986_WSi2}, OsSi$_2$\cite{Chang_JAP_1989_OsSi2}, and IrSi$_2$\cite{PhysRevB.79.104116}.  For many silicide compounds including PtSi (orthorhombic), NiSi (orthorhombic), and CoSi$_2$ (flourite), epitaxy occurs most readily on \textless111\textgreater ~substrates, though for lattice-matched fluorite structure compounds, epitaxy on \textless100\textgreater ~substrates is still possible under some preparation conditions.  For example, CoSi$_2$ and Si are both cubic with similar lattice parameters 5.3\AA~and 5.43\AA~respectively, and CoSi$_2$ can be grown epitaxially using high-temperature codeposition onto \textless100\textgreater ~substrates. 

In this work, we systematically study the thermal interface conductance of epitaxial and non-epitaxial interfaces of the metal-silicide TiSi$_2$, CoSi$_2$, NiSi and PtSi with silicon using time-domain thermoreflectance (TDTR) and compare the results to the most advanced available theories. 
\section{\label{sec:level2}Experimental}
\subsection{Epitaxial silicide growth}
TiSi$_2$, CoSi$_2$, NiSi and PtSi were fabricated under a wide range of conditions.  We studied films:  (i) on both Si(100) and Si(111) substrates,  (ii) using a wide range of Si substrate doping concentrations, (iii) using different surface cleaning methods, and (iv) two different growth techniques. The two different substrates orientations were used in order to generate different interfacial structures, since it is known that the rhombehedral compounds PtSi\cite{RN8,RN9,RN10,RN11} and NiSi\cite{RN12,RN13} films grow epitaxially on Si(111) surfaces, while these form polycrystalline structures on Si(100) surfaces.  All the silicides studied here were grown by thermally induced reactions of the pure metal:   Ti, Co, Ni, or Pt were deposited by RF-sputtering onto a Si substrate at 300K.  Samples were then annealed at high temperature (PtSi: 400 $^\circ$C; NiSi: 400$^\circ$C; TiSi$_2$: 750$^\circ$C; CoSi$_2$: 750$^\circ$C for 30 minutes) within the sputtering chamber to induce the reactive growth of the appropriate silicide layer ($\sim$110nm thick).  With the exception of TiSi$_2$ (C54 phase), the silicides here form epitaxial interfaces on Si(111) when grown by the thermal method.  None of the silicides form epitaxial interfaces when grown on Si(100) by this method.  In the case of CoSi$_2$ we also grew samples by co-sputtering of elemental Si and Co at 750$^\circ$C, which allowed the formation of epitaxial interfaces on Si(100) substrates, unlike the thermal method.  We pre-cleaned all Si wafers using acid piranha followed by either (1) an in-situ RF sputtering substrate bias cleaning, followed by a 750$^\circ$ substrate anneal or (2) an HF dip performed $\sim$30 sec prior to loadlocking the samples into the sputtering chamber.  The latter approach produced smoother final surfaces according XRR characterization.  Samples with substrate doping levels ranging from $n=1\times10^{19}$ cm$^3$ to $p=1\times10^{19}$ cm$^3$ were also created to explore electronic effects on the thermal interface conductance of metal-semiconductor junctions.
\begin{figure*}[htb]
\includegraphics[width=\textwidth]{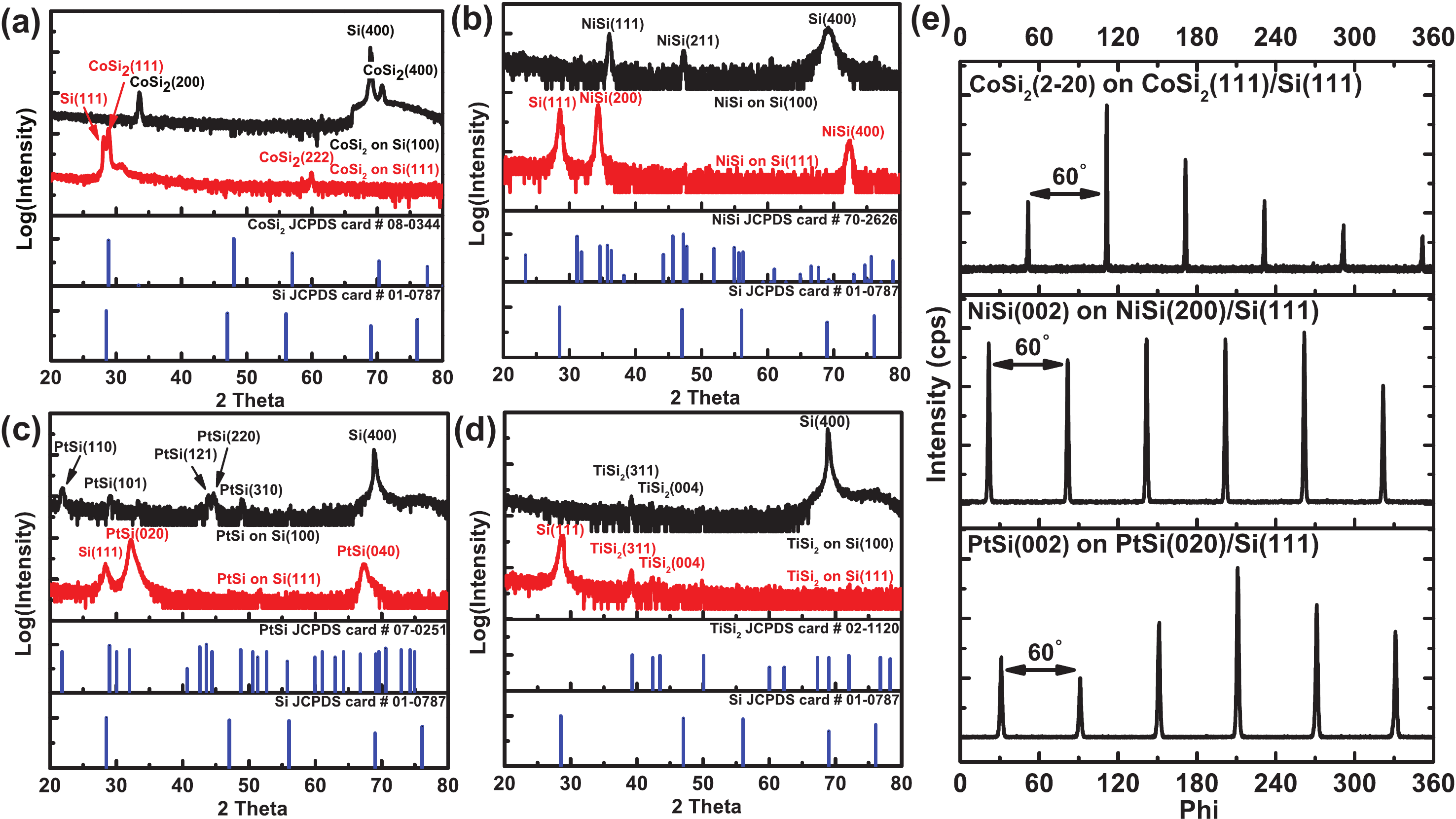}%
\caption{\label{fig:XRD}(a) XRD results of CoSi$_2$ on intrinsic Si(100) and Si(111) wafer; (b) XRD results of NiSi on intrinsic Si(100) and Si(111) wafer; (c) XRD results of PtSi on intrinsic Si(100) and Si(111) wafer; (d) XRD results of CoSi$_2$ on intrinsic Si(100) and Si(111) wafer; (e) XRD phi scan of the in-plane diffraction for PiSi(020)/Si(111), NiSi(200)/Si(111) and CoSi$_2$(111)/Si(111) samples.}
\end{figure*}
\begin{figure}
\includegraphics[width=0.40\textwidth]{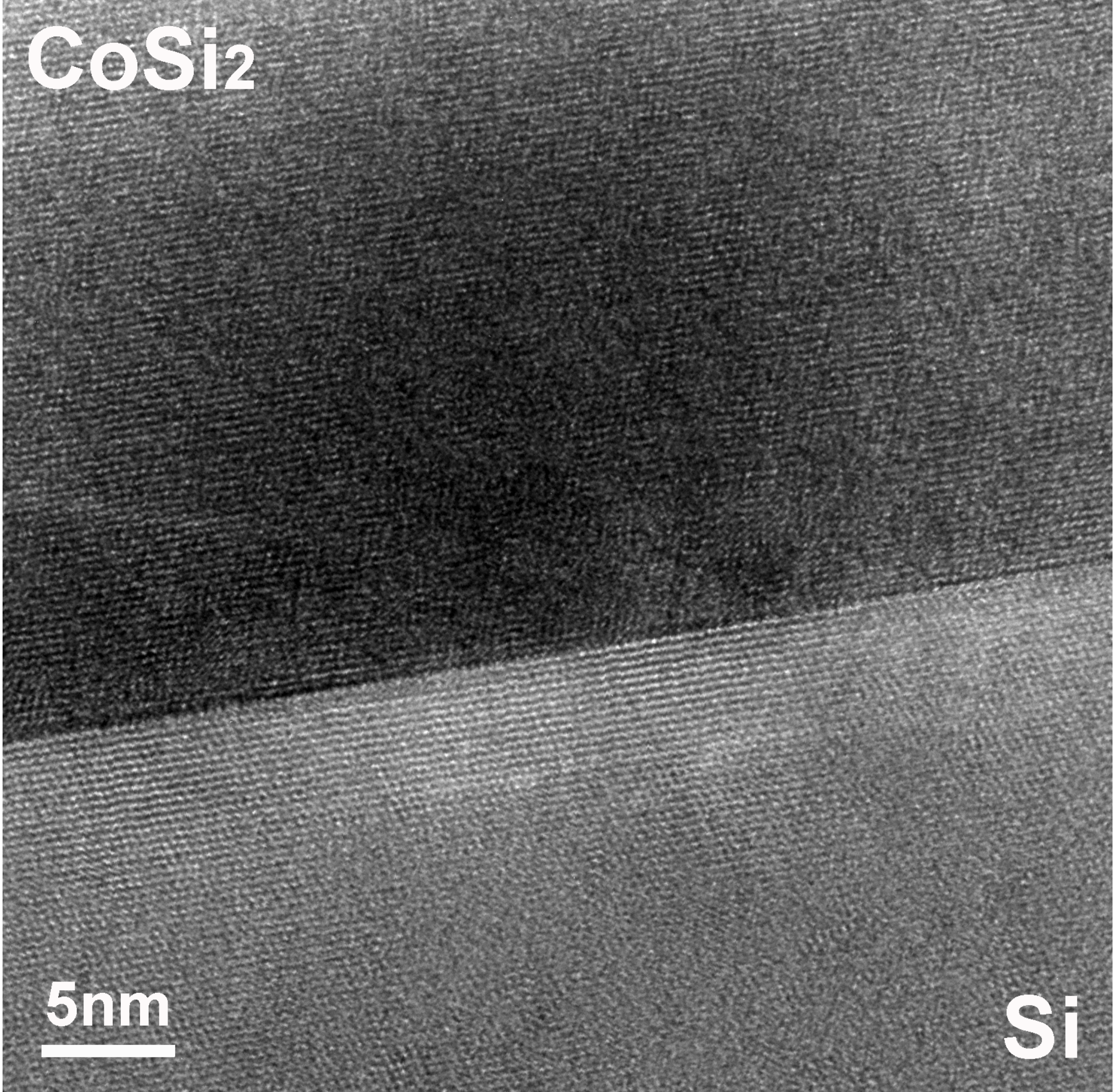}%
\caption{\label{fig:TEM} HRTEM of an epitaxial CoSi$_2$-Si interface.}
\end{figure}

X-ray diffraction (XRD) was used to characterize the structure of the films, and confirmed epitaxial growth in the cases of NiSi, PtSi and CoSi$_2$ on Si(111) substrates (regardless of which surface preparation was used) and also in the case of co-sputtered CoSi$_2$ on Si(100).  Figure 1 shows XRD $\omega$-2$\theta$ scans of different silicides grown on intrinsic Si(100) and Si(111) wafers, and the JCPDS index cards of the corresponding materials are given for comparison. No peaks from elemental Ti, Co, Ni or Pt are observed, indicating there is no unreacted metal detected in the films.  Scans for co-sputtered CoSi$_2$ on HF-pretreated Si(111) and Si(100)  show peaks at CoSi$_2$(111)/(222) and CoSi$_2$(200)/(400) respectively (Figure 1a). For NiSi on Si(111)  only NiSi(200)/(400) peaks at 34.4$^\circ$ and 72.5$^\circ$ respectively are observed (Figure 1b). In contrast, the NiSi on Si(100) shows primary peaks from NiSi(210) and Ni(211) diffraction planes (36.1$^\circ$ and 47.5$^\circ$), indicating oriented polycrystalline growth. The XRD patterns of PtSi/Si(111) and PtSi/Si(100) show similar behavior: PtSi films formed on Si(111) shows only PtSi(020)/PtSi(040) planes and the PtSi film grown on Si(100) exhibits a polycrystalline structure with almost random orientations.  For TiSi$_2$ on Si(111) and Si(100), both substrate orientations yield the same characteristic peaks corresponding to the C54 face-centered orthorhombic phase of TiSi$_2$ (the strongest of which are TiSi$_2$(311) and TiSi$_2$(004)), which is the same phase used in many microelectronics applications. We found no preferential growth direction in the case of TiSi$_2$ on Si(111) or Si(100). The growth of silicides on doped Si wafers was found to be identical to that on the intrinsic wafers.

XRD $\phi$-scans of the diffraction planes perpendicular to the sample surface (in-plane diffraction planes) were performed to confirm epitaxy of the silicides. Figure 1(e) shows the in-plane XRD $\phi$ scans of sample PtSi(020)/Si(111) , NiSi(200)/Si(111) and CoSi$_2$(111)/Si(111). For PtSi(020)/Si(111), the in-plane diffraction peak of PtSi(200) plane, corresponding to 2$\theta$=30.06$^\circ$, was taken while rotating the sample 360$^\circ$ with its out-of-plane axis. The pattern shows six-fold symmetry. While the crystal structure of PtSi is orthorhombic, the in-plane XRD $\phi$ scan shows a six-fold symmetry rather than two fold because there are 3 equivalent PtSi epitaxies conforming to the pseudo-hexagonal structure of the Si(111) surface.\cite{RN9} The same is expected to occur on orthorhombic NiSi on Si(111).\cite{RN14} The XRD $\phi$ scan of the in-plane diffraction peak of NiSi(020) plane with 2$\theta$=54.94$^\circ$ on the NiSi(200)/Si(111) sample shows a six-fold symmetry, which confirms a pseudo-hexagonal epitaxial growth of NiSi on Si(111). The XRD $\phi$ scan of CoSi$_2$(111)/Si(111) sample also indicates a six-fold symmetry of the in-plane diffraction peak of CoSi$_2$(2-20) plane with 2$\theta$=48.15$^\circ$. This result suggests the epitaxial growth of the CoSi$_2$(111) on Si(111). The in-plane lattice mismatch between the epitaxial silicide films and Si wafer are calculated to be $\approx$11\% for PtSi(020)/Si(111), $\approx$5\% for NiSi(200)/Si(111) and $\approx$1\% for CoSi$_2$(111)/Si(111).  Despite the large mismatch in the case of PtSi, the interface is known to form epitaxially by relieving strain using a undulating interface\cite{Kawarada}.

\subsection{Transport Characterization}
Thermal interface conductance and substrate thermal conductivity measurements were performed using time-domain thermoreflectance (TDTR).  Our system is based on the two-tint approach described by Kang and Cahill\cite{Kang_Cahill_twotint_RSI_2008}.  The measurement system and methods of data reduction have been described in detail previously.\cite{Kang_Cahill_twotint_RSI_2008,Cahill_TDTR_RSI_2004}  The time evolution of surface temperature is measured through temperature-dependent changes in the reflectivity, i.e., the thermoreflectance.  We analyze the ratio of in-phase $V_{in}$,  and out-of-phase $V_{out}$ variations in the intensity of the reflected probe beam at the modulation frequency (12.6 MHz unless otherwise specified) of the pump beam as a function of delay time between pump and probe.  The wavelength of the mode-locked Ti:sapphire laser is $\lambda=$785 nm and the $1/e^2$ radius of both focused beams is 25 $\mu$m with a repetition rate of 76 MHz. 

One unique aspect of this work is that we use the metal silicide itself as the metal transducer.  Unless otherwise noted, TDTR data reduction consisted of simultaneous non-linear least square extraction of the substrate thermal conductivity and thermal interface conductance between the silicide and silicon substrate. In order to perform data reduction, it is necessary to know the thickness of the silicide films, the heat capacity of all the layers, and the thermal conductivity of any layers for which data reduction is not being performed. The thickness of the silicides was determined by calibrating the thickness of pure metal deposited under the same conditions, measured by X-ray reflectivity, and using knowledge of the silicide lattice constant and stoichiometry. The characteristic light absorption length of the silicides ranged from 21-36nm (the details of this calculation are given in the Supporting Information), which is $\approx$3-fold larger compared to a traditional Al transducer. Therefore the silicide transducer layers were grown to be $\sim$110nm to ensure full absorption of the laser and to avoid anomalous signals at short time delays due to electron-hole pair modulations of the reflectivity. The absorption process was approximated using the bilayer technnique described by Cahill\cite{Cahill_TDTR_RSI_2004}, though the particular model used was not found to affect the experimental regression because the fit was performed at long time delays (300ps-3700ps) where details of the initial heat deposition profile no longer matter.  Most of the sensitivity to interface conductance occurs at the largest time delays where this is especially unimportant (details of the sensitivity analysis and errorbar estimation are given in the Supporting Information).  Heat capacities of all the films were determined by density functional perturbation theory (DFPT) through our own DFPT calculations. Results from the heat capacity calculations for CoSi$_2$, TiSi$_2$, NiSi and PtSi can be found in the Supporting Information.  The sheet resistance/electrical conductivity of the silicide films were measured at room temperature using an inline four point probe with regression to an I-V curve.  The resulting electrical resistivities were used to estimate the electronic component to thermal conductivity using the Wiedemann-Franz law, assuming the degenerate Lorenz number $L_0$=2.44x10$^{-8}$ W$\Omega$/K$^2$.  The results are given in Table~\ref{tab:silicide_resistivity}.  The thermal conductivity of silicides was high enough to yield good sensitivity to thermal interface conductance, and the electronic thermal conductivity was verified to be a good approximation of the total thermal conductivity with TDTR using observations at time-scales below 400 ps, where sensitivity of the signal to the metal transducer’s thermal conductivity is strongest and most independent.  The temperature dependent electrical resistivity of CoSi$_2$, TiSi$_2$ and NiSi used in this work were estimated by using the literature-reported temperature coefficient of resistance\cite{CoSi2_temp, TiSi2_temp, NiSi_temp} combined with our measured room temperature electrical resistivity values (coefficients and equations given in Supporting Information). Note that errorbars associated with the extracted thermal interface conductance are found to be much less sensitive to the transducer thermal conductivity than the extracted substrate thermal conductivity.  For example, for CoSi$_2$ at 300K a 10\% uncertainty in silicide thermal conductivity corresponds to an errorbar of 4.6\% in the extracted interface conductance and 13.6\% in substrate thermal conductivity. For the case of PtSi, there are no previously reported temperature coefficients of resistance. The PtSi/Si temperature dependent TDTR data were therefore analyzed by including PtSi silicide thermal conductivity as a third fitting parameter. At room temperature, the thermal interface conductance value of PtSi/Si obtained this way was within 5\% of that obtained in Fig.~\ref{fig:ThermalIntCond} using the measured thermal conductivity.

In the cases of PtSi, NiSi, CoSi$_2$, TiSi$_2$ (C54), we found that the temperature dependence of reflectivity at 300K is comparable to the best previously reported materials\cite{YuxinWang_thermoreflectance_coeff_2010} at $\lambda$=785 nm.  The thermoreflectance coefficients of CoSi$_2$ films had a positive value, while TiSi$_2$, NiSi and PtSi exhibited negative values.  In the case of CoSi$_2$, the thermoreflectance coefficient was found to switch signs near 600K, allowing substrate thermoreflectance effects to become experimentally visible and thus complicating the data analysis.  For this reason we restrict our experimental results for CoSi$_2$ to room temperature and below.  This effect was not observed in PtSi, NiSi and TiSi$_2$, which allowed measurements from 77K-700K for these materials. 
 \begin{table}
 \caption{\label{tab:silicide_resistivity}Silicide properties at 300K}
 \begin{ruledtabular}
 \begin{tabular}{cccc}
 Silicide & $\rho$ ($\mu\Omega$-cm) & $\kappa_e$ (W/m-K) & $C_V$ (10$^6$ J/m$^3$-K)\\
 \hline
CoSi$_2$ & 16 & 44 (calc) & 2.74\\
TiSi$_2$ & 19 & 38 (calc) & 2.52\\
NiSi & 20 & 36 (calc) & 2.99\\
PtSi & 40 & 18 (calc) & 2.49\\
 \end{tabular}
 \end{ruledtabular}
 \end{table}
\section{Modeling}
\begin{figure}
\includegraphics[width=0.40\textwidth]{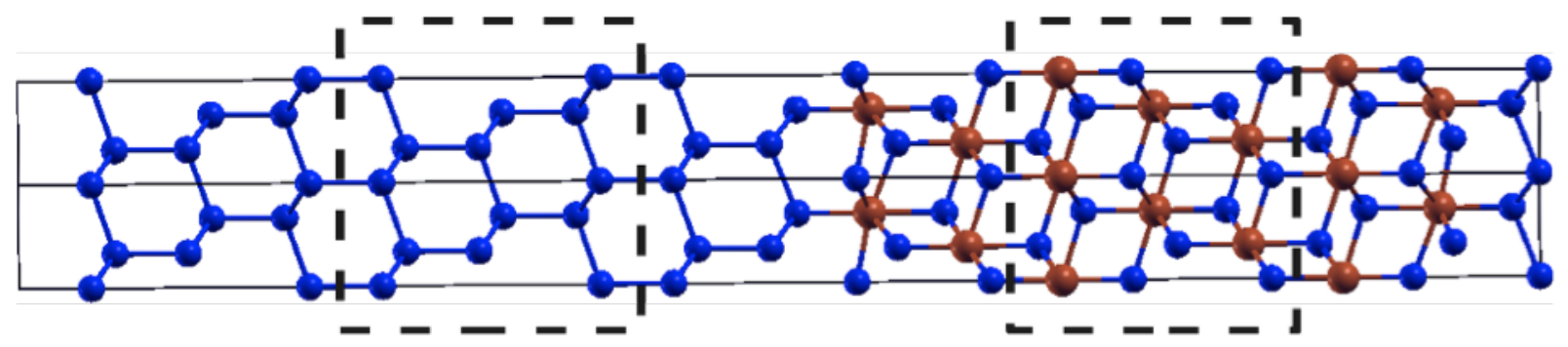}%
\caption{\label{fig:supercell} Schematic of the Si-CoSi$_2$ interface supercell with a 8B interfacial atomic configuration. The dashed rectangular boxes indicate the unit cells of bulk Si and bulk strained CoSi$_2$.}
\end{figure}

Two forms of phonon transport modeling have been used, here:  (1) interface thermal conductance calculations from full-phonon-dispersion diffuse mismatch modeling\cite{RN18}, using  phonon dispersions obtained from DFPT, and  (2) atomistic Green's function simulations employing density functional theory (DFT) to calculate interatomic force constants, including the effect of bonding changes near the interface.    First-principles calculations for phonon dispersion were performed in the case of CoSi$_2$ and Si under the density functional theory (DFT) framework using Quantum Espresso\cite{RN25}, with a planewave basis set. The exchange correlation energy was approximated under the generalized gradient approximation (GGA) using the Perdew-Burke-Ernzerhof (PBE) functional form. Rappe-Rabe-Kaxiras-Joannopoulos (RRKJ) ultrasoft pseudopotentials were used for both Si and Co atoms. The relaxed bulk lattice constant of Si and CoSi$_2$ were found to be 5.46 \AA ~and 5.36 \AA ~which compares well with the experimental lattice constants of 5.43 \AA ~(Si) and 5.36 \AA ~(CoSi$_2$). While the density functional perturbation theory (DFPT) calculations on bulk Si and bulk CoSi$_2$ provide the bulk inter-atomic force constants (IFCs), the AGF method also requires as input the interfacial force constants at the Si-CoSi$_2$ interface. We perform DFT/DFPT calculations on a Si (111)-CoSi$_2$ (111) interface supercell shown in Figure~\ref{fig:supercell}. This interface supercell corresponds to the 8B interfacial atomic configuration that has been identified to have the lowest interfacial energy in prior first-principles calculations of the Si-CoSi$_2$ interface,\cite{RN26} and was also experimentally observed using TEM (Fig. \ref{fig:TEM}).  A tensile strain of 1.8\% is imposed on CoSi$_2$ to match its lattice with Si. All the atomic positions and the lattice constant along the c-direction (heat transport direction) are relaxed for the interface supercell while the in-plane lattice constants of the interface supercell are fixed to that of bulk Si. Because the AGF simulations are performed on an interface between Si and strained CoSi$_2$, the bulk phonon dispersion and IFCs of strained CoSi$_2$ are also determined using a separate DFT/DFPT calculation. The unit cell in DFT calculations of bulk strained CoSi$_2$ corresponds to a 9 atom unit cell as shown the dashed box in Figure~\ref{fig:supercell}. We also use a 6 atom unit cell for calculations on bulk Si with one of the lattice vectors aligned along the [111] direction (see Figure~\ref{fig:supercell}). Table~\ref{tab:DFTdetails} shows the k-point grids and the cutoff energies used in DFT calculations of the bulk and interface structures. The phonon dynamical matrices are computed using DFPT on a Monkhorst-Pack q-point grid, and the phonon dispersion at arbitrary q-points are obtained using Fourier interpolation. 
 \begin{table}
 \caption{\label{tab:DFTdetails} DFPT Details for CoSi$_2$ / Si interface modeling}
 \begin{ruledtabular}
 \begin{tabular}{ccccc}
Parameter & Bulk & Bulk & Bulk  & Si-CoSi$_2$\\
 & Si & unstrained  & strained & supercell\\
 & & CoSi$_2$ & CoSi$_2$ & \\
 \hline
Kinetic energy & 680 & 820 & 820 & 820\\
cutoff \\
Charge density & 6800 & 8200 & 8200 & 8200\\
cutoff \\
Electron k-point & 12$\times$12$\times$9 & 14$\times$14$\times$14 & 16$\times$16$\times$12 & 16$\times$16$\times$1 \\
grid\\
Phonon q-point & 4$\times$4$\times$3 & 4$\times$4$\times$4 & 4$\times$4$\times$3 & 4$\times$4$\times$1\\
grid\\
 \end{tabular}
 \end{ruledtabular}
 \end{table}

The bulk phonon dispersions of Si and CoSi$_2$ are used to obtain predictions for the thermal interface conductance using the diffuse mismatch model, and an upper limit for the elastic interface conductance is obtained from the radiation limit. Our DMM predictions use the exact phonon dispersion of bulk Si and bulk CoSi$_2$ as opposed to a Debye approximation that is commonly used in the literature. The procedure for full-dispersion DMM is described in Ref.~\onlinecite{RN18}, and the details of the radiation limit are provided in Ref.~\onlinecite{RN19}. 
The AGF method uses harmonic inter-atomic force constants (IFCs) to determine the phonon transmission function that is then used in the Landauer approach to determine the thermal interface conductance. 

An important development in the present work is the prediction of cross-interface force constants directly from DFPT calculations on a Si (111)-CoSi$_2$ (111) interface supercell.  Our approach is a significant improvement  in comparison with common heuristic approximations such as averaging of bulk force constants to obtain interface force constants. Such rigorous predictions of interface bonding strength is important since the phonon transmission function is strongly sensitive to the strength and nature of interfacial bonding.  Since the AGF simulations consider strained CoSi$_2$, we also perform DMM calculations using the bulk phonon dispersion of strained CoSi$_2$. AGF simulations model specular reflection and transmission of phonons at the interface while the DMM assumes that the interface destroys all phase and direction information for the phonons incident on the interface. Hence, the AGF and DMM approaches are expected to represent perfectly smooth and rough interfaces respectively. 

In the case of PtSi, the interatomic force constants (IFC) and phonon frequencies were calculated using DFT and PBE revised for solids (PBEsol) \cite{PBEsol} as implemented in the Vienna {\em Ab initio} simulation package (VASP) \cite{VASP-1,VASP-2}. Projector augmented wave (PAW) potentials \cite{PAW} are used to describe the interaction between the valence electrons and the ion cores, and an energy cutoff of 500 eV was used for plane wave expansion. We first determined the lattice parameter of PtSi by using a primitive cell with 8 atoms, in the orthorhombic structure, and a 8×8×8 Monkhorst-Pack mesh for integrations over the Brillouin zone. The calculated lattice parameters of $a$=3.60, $b$=5.59 and $c$=5.92\AA~are in good agreement with experimental values of 3.59, 5.57, and 5.91\AA\cite{PtSi}. The second order IFCs were calculated using the Phonopy code \cite{phonopy} with a supercell of 64 atoms, which is a 2$\times$2$\times$2 repetition of the 8-atom primitive cell, and a 2$\times$2$\times$2 mesh of special $k$-points were used in these calculations. The IFCs matrix in real space was converted to dynamic matrix in reciprocal space by Fourier transforms. The phonon frequencies were then obtained by solving the eigenvalue problem of the dynamic matrix for phonon $q$ vectors in the Brillouin zone sampled by a 50$\times$50$\times$50 mesh in ShengBTE code~\cite{ShengBTE}. DMM modeling for NiSi and TiSi$_2$ were done in an analogous way, but using their corresponding lattice structures.
\section{Results and Discussion}
\subsection{Thermal Interface Conductance of Epitaxial and non-epitaxial silicide-silicon interfaces}
\begin{figure}[!htb]
\includegraphics[width=0.45\textwidth]{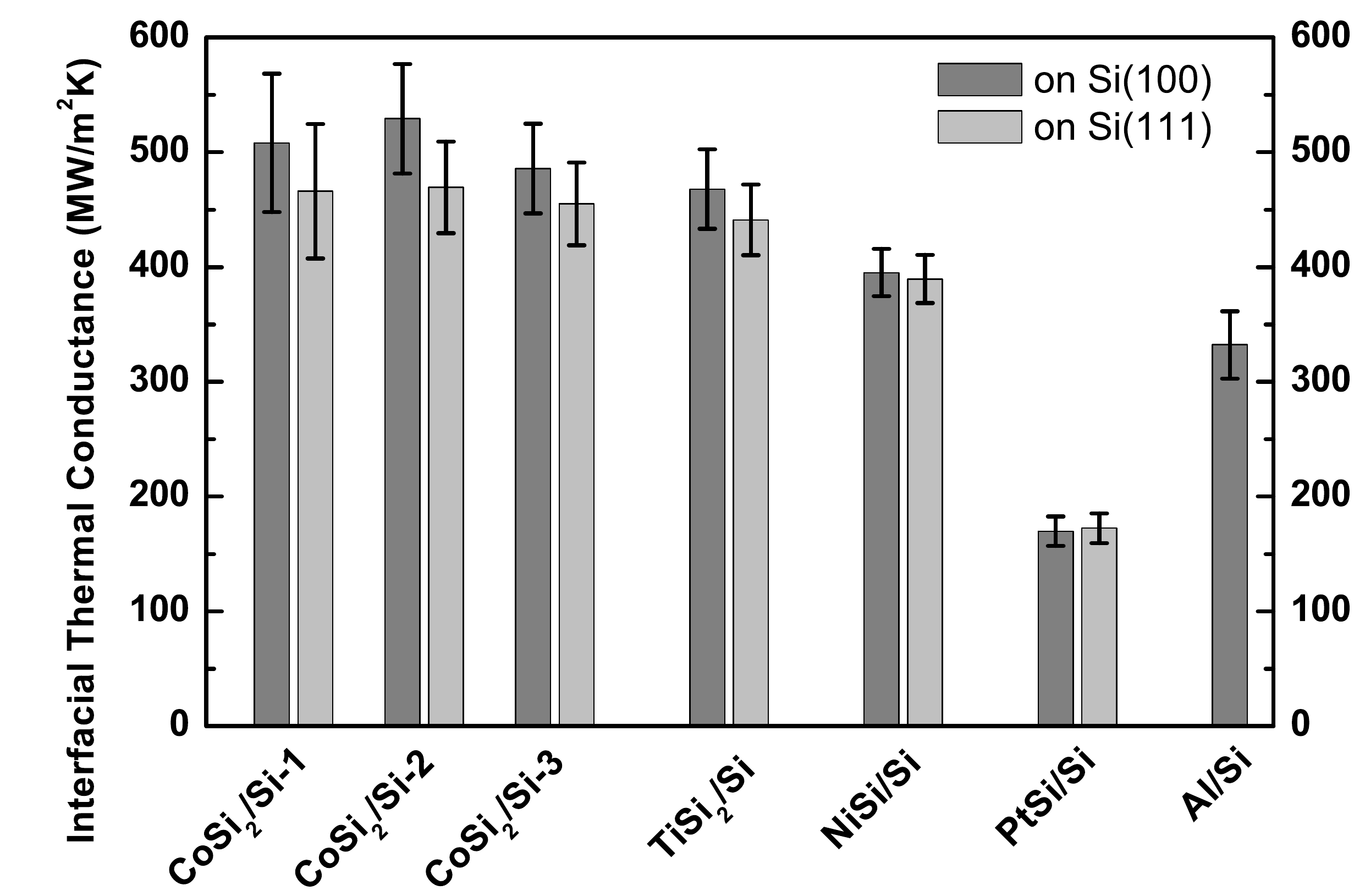}%
\caption{\label{fig:ThermalIntCond} The interfacial thermal conductance of CoSi$_2$, TiSi$_2$, NiSi and PtSi on intrinsic Si(100) and Si(111) wafer. CoSi$_2$/Si-1, CoSi$_2$/Si-2 and CoSi$_2$/Si-3 represent samples made under different conditions. CoSi$_2$/Si-1: HF treated wafer + films deposited by co-sputtering, CoSi$_2$/Si-2: HF treated wafer + films made by reactive growth method, CoSi$_2$/Si-3: RF bias treated wafer + films made by reactive growth method. For TiSi$_2$, NiSi and PtSi films are made by reactive growth method on RF-bias cleaned substrates. The interfacial thermal conductance of Al/Si(100) is also attached as a reference.}
\end{figure}

The results for room temperature thermal interface conductance of the silicide/Si interfaces on intrinsic silicon substrates are given in Figure~\ref{fig:ThermalIntCond}. The thermal conductance of the CoSi$_2$/Si and TiSi$_2$/Si were both near 480 MW/m$^2$-K, similar to a recent report for an epitaxial NiSi$_2$/Si interface,\cite{RN7} but greatly exceeding the highest interfacial thermal conductance for all other previously measured interfaces on silicon, including HF dipped Al/Si interfaces.  To within the experimental uncertainty, there was no difference between the measured values of interfacial conductance formed on Si(100) vs Si(111).  In other words, the epitaxial interfaces show nearly the same thermal interface conductance as the non-epitaxial interfaces in all cases.  However, it appears that if the sources of error in TDTR are systematic (as they are usually are), the non-epitaxial interfaces may even have marginally larger thermal interface conductance than the epitaxial interfaces.  Furthermore, in the case of CoSi$_2$/Si interfaces, we found no dependence of the thermal interface conductance on the surface preparation method (in situ RF-bias cleaning vs. HF dipped) or the method of silicide formation (reactive method vs. co-sputtering).  The interfacial thermal conductance of silicide interfaces is thus found to be quite robust and high so long as the wafer surface is cleaned before the silicide formation.
NiSi/Si interfacial thermal conductance is also relatively high, $G_{\textrm{NiSi/Si}}\approx400$ MW/m$^2$-K. PtSi has much larger acoustic and phonon density of states contrast with Si compared to the other materials, and as expected its thermal interface conductance is substantially smaller than the other silicides studied, with $G_{\textrm{PtSi/Si}}\approx170$ MW/m$^2$-K.

While the observation may seem surprising that the thermal interface conductance of epitaxial silicides is essentially identical to those of non-epitaxial silicides, it is not without precedent.  Similar results were reported for epitaxial TiN(001)/MgO(001), TiN(111)/ MgO(111) and TiN(111)/Al$_2$O$_3$(0001) interfaces,\cite{RN6} where it was found that, despite significant differences in lattice mismatch (8\% when comparing O-O and N-N distances in the case of TiN/Al$_2$O$_3$(0001)) and the presence of stacking faults in the case of both TiN(111)/Al$_2$O$_3$(0001) and  TiN(111)/MgO(111), all the interfaces showed the same interface thermal conductance.  In that work the authors cite two possible reasons why this might be the case:  (1) all samples undergo strong phonon scattering at the interface (including the seemingly perfectly structured ones) and therefore, all samples satisfy the assumptions of the diffuse mismatch model or (2) the interface disorder in all samples (including the more disordered ones) are weak and the transmission coefficient is always close to unity.  We should note that the authors explicitly calculated the DMM for these cases and did not find good agreement.   However, they implemented a relatively crude approach to perform the diffuse mismatch model calculations (Debye model).  It is well established now that using full phonon dispersions produces substantially different DMM predictions under most circumstances.  In addition, the authors utilized lattice dynamics (LD) calculations to predict the results for a perfect epitaxial interface.  However, they did not consider local changes in bonding characteristics near the interface, which may also have been important. The DMM and AGF calculations here take these into account.  Also, by comparing the experimental data to the calculated  radiation limit (the maximum interface conductance consistent with detailed balance in the elastic limit) using full-dispersion relations, we are able to test the hypothesis (2) directly.  In all cases, we find that silicide-silicon interfaces are not close to the radiation limit and thus the transmission coefficients are not close to unity (or rather the maximum allowable) for all modes.

\begin{figure*}[!htb]
\includegraphics[width=\textwidth]{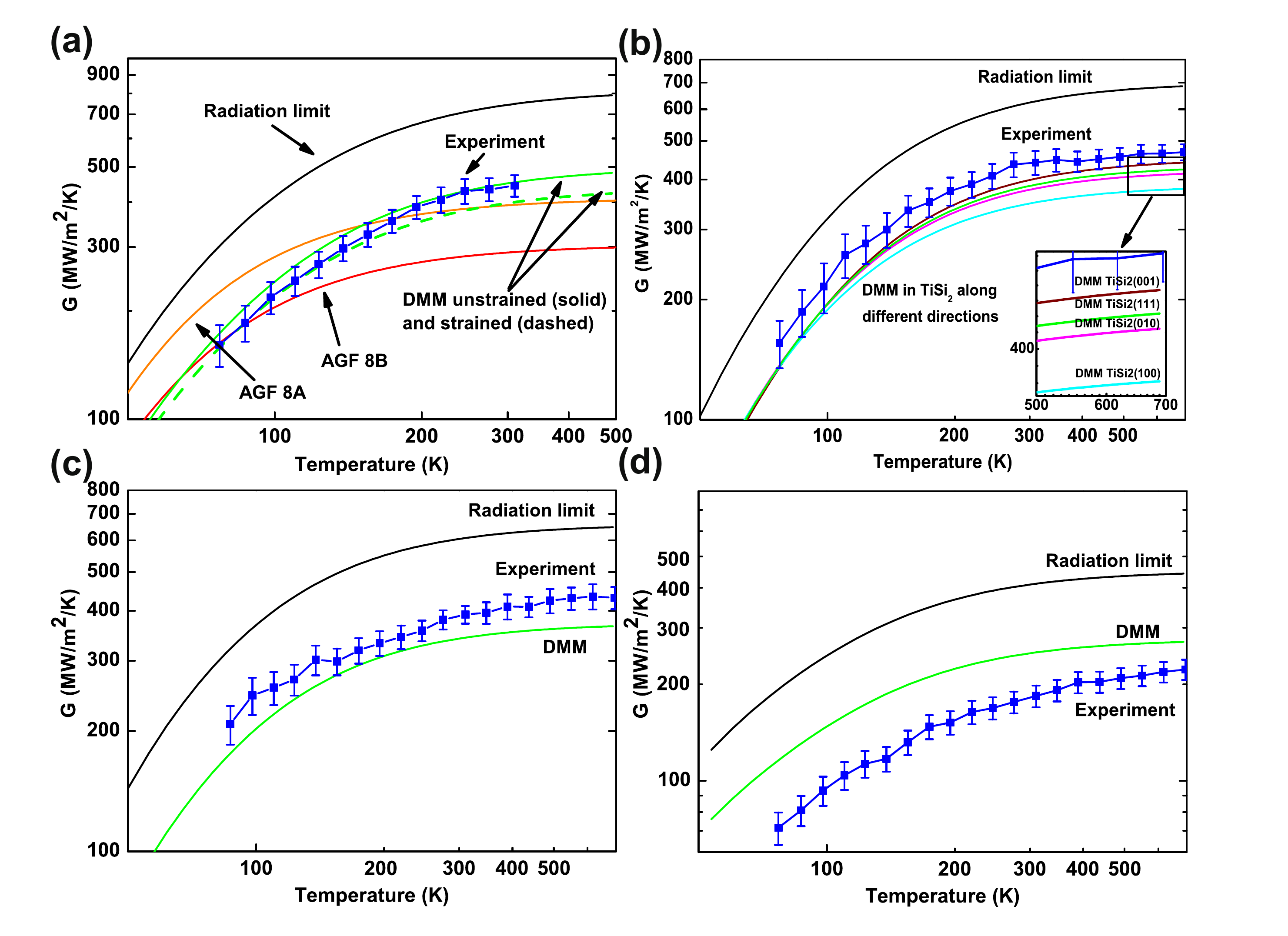}
\caption{\label{fig:Compare2Theory} (a)Modeling results for CoSi$_2$(111)-Si(111) interfaces using various models:  the full-dispersion diffuse mismatch model (green), the atomistic Green’s function method for interface of 8B(red) and 8A(orange) and the radiation limit (black). Experimental data at room temperature is shown for comparison (blue squares); (b)Comparison between experimental thermal interface conductance of TiSi$_2$-Si(111) interface (blue squares), the full-dispersion DMM calculation of TiSi$_2$(001), TiSi$_2$(010), TiSi$_2$(100), TiSi$_2$(111)-Si(111) and the radiation limit (black); (c)Comparison between experimental thermal interface conductance of an epitaxial NiSi(200)-Si(111) interface (blue squares), the full-dispersion DMM calculation (green) and the radiation limit (black);(d)Comparison between experimental thermal interface conductance of an epitaxial PtSi(020)-Si(111) interface (blue squares), the full-dispersion DMM calculation (green) and the radiation limit (black).}
\end{figure*}

Figure~\ref{fig:Compare2Theory}(a) shows a comparison between temperature-dependent  CoSi$_2$ experiments and our DMM and AGF calculations.\cite{sadasivam2016thermal} Since the experimental Si (111)-CoSi$_2$ (111) interfaces considered here are epitaxial with submonolayer interfacial roughness it would be reasonable to expect the AGF method to be applicable. We observe, however, that the experimental value of thermal interface conductance ($\approx$500 MW/m$^2$-K) exceeds the AGF prediction by more than 50\%. At room temperature a full dispersion DMM not accounting for interfacial strain nearly accounts for the data at high temperature. However, employing a modified DMM that incorporates the effect of strain yields slightly worse agreement ($\approx$10\%).

Given the lack of dependence of the interface conductance upon interface structure (i.e. epitaxial vs. not in Figure ~\ref{fig:ThermalIntCond}) and the reasonable agreement between experiments and the DMM model in Figure ~\ref{fig:Compare2Theory}, it may be tempting to assume that interfaces really act as diffusely to phonons, even for epitaxial interfaces.  However, we would caution that the observed high temperature discrepancies could also arise without the diffuse assumption, through a combination of inelastic interfacial processes and inter- and intra-material electron-phonon coupling.  Neither these processes is included current the model. Electron-phonon coupling within the metal provides a series resistance to the phonon-phonon interface resistance, while cross-interface electron-phonon coupling provides a parallel pathway for coupling between the primary energy carriers of metal and the semiconductor. Sadasivam et al.\cite{RN22} performed first-principles calculations of electron-phonon coupling near a C49 TiSi$_2$-Si interface and found that the coupling of electrons with joint or interfacial phonon modes can potentially produce a conductance similar to the phonon-phonon interfacial conductance (note: in the present paper, we obtained the C54 phase of TiSi$_2$ which is the lower resistivity phase and is more commonly used for semiconductor applications).  Inelastic phonon scattering has been identified as an important transport mechanism for material combinations with a large acoustic mismatch such as Pb and diamond \cite{RN20,RN21}. In the case of CoSi$_2$ of Si(111), we show in a forthcoming publication that the high temperature behavior of interface conductance can be matched quite well by invoking these mechanisms\cite{sadasivam2016thermal}.  It remains unclear, however, whether these mechanisms are insensitive to interfacial structure.

\subsection{The effect of substrate carrier concentration}
\begin{figure}[!htb]
\includegraphics[width=0.5\textwidth]{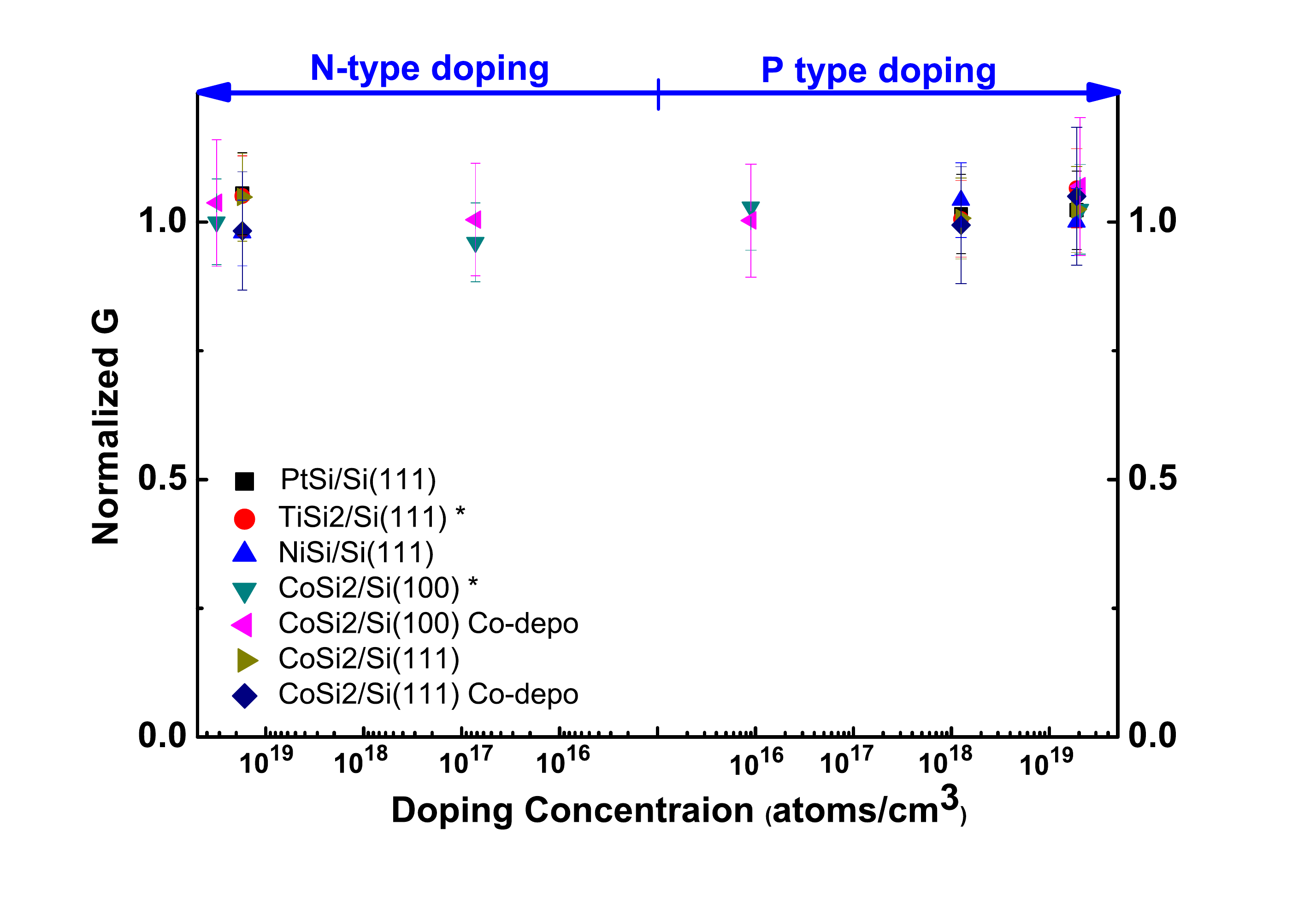}%
\caption{\label{fig:DopingDependence}The substrate doping effects on the interfacial thermal conductance. The thermal conductance values are normalized with their correspondent undoped values. This includes the interfacial thermal conductance of CoSi$_2$/Si(111) and CoSi$_2$/Si(100) made by co-deposition, CoSi$_2$/Si(111), CoSi$_2$/Si(100), TiSi$_2$/Si(111), NiSi/Si(111) and PtSi/Si(111) made by reactive growth. * indicates that the interface is not epitaxial. (For original data please refer to figure 5 of the supplemental document)}
\end{figure}

While we are not aware of any experimental methods capable of isolating the cross-interface electron-phonon coupling component of thermal conductance from phonon-phonon transport across the interface, we have studied the effect of doping the silicon wafer on the interfacial thermal conductance at room temperature (Figure~\ref{fig:DopingDependence}).  To our knowledge, there are no previous reports of the substrate carrier concentration dependence of thermal interface conductance for any metal-semiconductor interface.  If the thermal interface conductance is phonon-dominated, the doping level of the substrate would not be expected to have any effect on the thermal interface conductance, perhaps justifying the lack of existing experiments.  On the other hand, if cross-interface electron-phonon coupling is dependent on either band-bending, through Schottky barrier height and depth, or electronic screening, which depends on carrier concentration through the screening length, then we reason that substantial changes to the substrate carrier concentration could affect the electron-phonon coupling component of the thermal interface conductance. 

We have studied a wide range of carrier concentrations from degenerate n- to degenerate p-type doping and have found that the thermal interface conductance of the silicides do not show any carrier concentration dependence.  In fact, the interface conductance is remarkably consistent and stable against changes to both doping level/type and substrate/film orientation as shown in Figure 5.  As has been previously reported, however, we do find that the substrate thermal conductivity is appreciably reduced for degenerate levels of either p- or n- type dopants (red squares in Figure~\ref{fig:DopingDependence}).  Note that the extracted substrate conductivities were consistent across the various silicides, which gives additional confidence in the values of heat capacity for the silicides used to extract TDTR data. The decrease in the thermal conductivity of the silicon wafer with increase of the doping concentration (Figure 5 in supplemental document) is due to the phonon scattering by the impurity atoms and electrons (or holes)\cite{RN23}. The measured thermal conductivity of the silicon wafer is consistent with past measurements of intrinsic/doped Si performed by TDTR.\cite{RN24}
\section{Conclusions}
In summary, the thermal interface conductance of CoSi$_2$, NiSi and PtSi, TiSi$_2$, and Al films with silicon were measured using a series of clean and well-characterized interfaces using TDTR.  A few of these (CoSi$_2$, TiSi$_2$ and NiSi) are demonstrated to be extremely good thermal interface conductance materials for Si, and are some of the highest ever measured for a metal-semiconductor interface.  Interestingly, we find that the interfacial thermal conductance is not dependent on whether the interface is epitaxially grown or what the substrate orientation is. Above 100K a coherent AGF approach significantly underpredicts interface conductance in the case of CoSi$_2$ suggesting that energy transport does not occur purely by coherent transmission of phonons, even for epitaxial interfaces.  A full-dispersion diffuse mismatch model closely predicts the experimentally observed interface conductances for CoSi$_2$, NiSi, and TiSi$_2$ interfaces, while it remains an open question whether inelastic scattering, cross-interfacial electron-phonon coupling, or other mechanisms could also account for the high temperature behavior\cite{sadasivam2016thermal}.  The effect of degenerate semiconductor dopant concentration on metal-semiconductor thermal interface conductance was also investigated with the result that we have found no dependencies of the thermal interface conductances up to (n-type or p-type) $\approx 1\times10^{19}$ cm$^{-3}$, indicating that there is no significant direct electronic transport and no transport effects which depend on long-range metal-semiconductor band alignment.
\section{Acknowledgements}
SS acknowledges financial support from the Office of Naval Research (Award No: N000141211006) and Drs. Helen and Marvin Adelberg fellowship from the School of Mechanical Engineering at Purdue University. The authors also acknowledge Dr.~Rosa Diaz Rivas who performed TEM imaging of the Si-CoSi$_2$ interface. 


\begin{thebibliography}{58}%
\makeatletter
\providecommand \@ifxundefined [1]{%
 \@ifx{#1\undefined}
}%
\providecommand \@ifnum [1]{%
 \ifnum #1\expandafter \@firstoftwo
 \else \expandafter \@secondoftwo
 \fi
}%
\providecommand \@ifx [1]{%
 \ifx #1\expandafter \@firstoftwo
 \else \expandafter \@secondoftwo
 \fi
}%
\providecommand \natexlab [1]{#1}%
\providecommand \enquote  [1]{``#1''}%
\providecommand \bibnamefont  [1]{#1}%
\providecommand \bibfnamefont [1]{#1}%
\providecommand \citenamefont [1]{#1}%
\providecommand \href@noop [0]{\@secondoftwo}%
\providecommand \href [0]{\begingroup \@sanitize@url \@href}%
\providecommand \@href[1]{\@@startlink{#1}\@@href}%
\providecommand \@@href[1]{\endgroup#1\@@endlink}%
\providecommand \@sanitize@url [0]{\catcode `\\12\catcode `\$12\catcode
  `\&12\catcode `\#12\catcode `\^12\catcode `\_12\catcode `\%12\relax}%
\providecommand \@@startlink[1]{}%
\providecommand \@@endlink[0]{}%
\providecommand \url  [0]{\begingroup\@sanitize@url \@url }%
\providecommand \@url [1]{\endgroup\@href {#1}{\urlprefix }}%
\providecommand \urlprefix  [0]{URL }%
\providecommand \Eprint [0]{\href }%
\providecommand \doibase [0]{http://dx.doi.org/}%
\providecommand \selectlanguage [0]{\@gobble}%
\providecommand \bibinfo  [0]{\@secondoftwo}%
\providecommand \bibfield  [0]{\@secondoftwo}%
\providecommand \translation [1]{[#1]}%
\providecommand \BibitemOpen [0]{}%
\providecommand \bibitemStop [0]{}%
\providecommand \bibitemNoStop [0]{.\EOS\space}%
\providecommand \EOS [0]{\spacefactor3000\relax}%
\providecommand \BibitemShut  [1]{\csname bibitem#1\endcsname}%
\let\auto@bib@innerbib\@empty
\bibitem [{\citenamefont {L{\'e}onard}\ and\ \citenamefont
  {Talin}(2011)}]{RN1}%
  \BibitemOpen
  \bibfield  {author} {\bibinfo {author} {\bibfnamefont {F.}~\bibnamefont
  {L{\'e}onard}}\ and\ \bibinfo {author} {\bibfnamefont {A.~A.}\ \bibnamefont
  {Talin}},\ }\href {\doibase 10.1038/nnano.2011.196 10.1038/NNANO.2011.196}
  {\bibfield  {journal} {\bibinfo  {journal} {Nature Nanotechnology}\ }\textbf
  {\bibinfo {volume} {6}} (\bibinfo {year} {2011}),\ 10.1038/nnano.2011.196
  10.1038/NNANO.2011.196}\BibitemShut {NoStop}%
\bibitem [{\citenamefont {Vilkov}\ \emph {et~al.}(2013)\citenamefont {Vilkov},
  \citenamefont {Fedorov}, \citenamefont {Usachov}, \citenamefont {Yashina},
  \citenamefont {Generalov}, \citenamefont {Borygina}, \citenamefont
  {Verbitskiy}, \citenamefont {Gruneis},\ and\ \citenamefont {Vyalikh}}]{RN2}%
  \BibitemOpen
  \bibfield  {author} {\bibinfo {author} {\bibfnamefont {O.}~\bibnamefont
  {Vilkov}}, \bibinfo {author} {\bibfnamefont {A.}~\bibnamefont {Fedorov}},
  \bibinfo {author} {\bibfnamefont {D.}~\bibnamefont {Usachov}}, \bibinfo
  {author} {\bibfnamefont {L.~V.}\ \bibnamefont {Yashina}}, \bibinfo {author}
  {\bibfnamefont {A.~V.}\ \bibnamefont {Generalov}}, \bibinfo {author}
  {\bibfnamefont {K.}~\bibnamefont {Borygina}}, \bibinfo {author}
  {\bibfnamefont {N.~I.}\ \bibnamefont {Verbitskiy}}, \bibinfo {author}
  {\bibfnamefont {A.}~\bibnamefont {Gruneis}}, \ and\ \bibinfo {author}
  {\bibfnamefont {D.~V.}\ \bibnamefont {Vyalikh}},\ }\href {\doibase
  10.1038/srep02168} {\bibfield  {journal} {\bibinfo  {journal} {Sci Rep}\
  }\textbf {\bibinfo {volume} {3}},\ \bibinfo {pages} {2168} (\bibinfo {year}
  {2013})}\BibitemShut {NoStop}%
\bibitem [{\citenamefont {Lee}\ and\ \citenamefont {Cahill}(1997)}]{RN3}%
  \BibitemOpen
  \bibfield  {author} {\bibinfo {author} {\bibfnamefont {S.~M.}\ \bibnamefont
  {Lee}}\ and\ \bibinfo {author} {\bibfnamefont {D.~G.}\ \bibnamefont
  {Cahill}},\ }\href {\doibase 10.1063/1.363923} {\bibfield  {journal}
  {\bibinfo  {journal} {Journal of Applied Physics}\ }\textbf {\bibinfo
  {volume} {81}},\ \bibinfo {pages} {2590} (\bibinfo {year}
  {1997})}\BibitemShut {NoStop}%
\bibitem [{\citenamefont {Nan}\ \emph {et~al.}(1997)\citenamefont {Nan},
  \citenamefont {Birringer}, \citenamefont {Clarke},\ and\ \citenamefont
  {Gleiter}}]{RN4}%
  \BibitemOpen
  \bibfield  {author} {\bibinfo {author} {\bibfnamefont {C.-W.}\ \bibnamefont
  {Nan}}, \bibinfo {author} {\bibfnamefont {R.}~\bibnamefont {Birringer}},
  \bibinfo {author} {\bibfnamefont {D.~R.}\ \bibnamefont {Clarke}}, \ and\
  \bibinfo {author} {\bibfnamefont {H.}~\bibnamefont {Gleiter}},\ }\href
  {\doibase 10.1063/1.365209} {\bibfield  {journal} {\bibinfo  {journal}
  {Journal of Applied Physics}\ }\textbf {\bibinfo {volume} {81}},\ \bibinfo
  {pages} {6692} (\bibinfo {year} {1997})}\BibitemShut {NoStop}%
\bibitem [{\citenamefont {Goodson}\ and\ \citenamefont {Ju}(1999)}]{RN5}%
  \BibitemOpen
  \bibfield  {author} {\bibinfo {author} {\bibfnamefont {K.~E.}\ \bibnamefont
  {Goodson}}\ and\ \bibinfo {author} {\bibfnamefont {Y.~S.}\ \bibnamefont
  {Ju}},\ }\href@noop {} {\bibfield  {journal} {\bibinfo  {journal} {Annual
  Review of Materials Science}\ }\textbf {\bibinfo {volume} {29}},\ \bibinfo
  {pages} {261} (\bibinfo {year} {1999})}\BibitemShut {NoStop}%
\bibitem [{\citenamefont {Stoner}\ and\ \citenamefont
  {Maris}(1993{\natexlab{a}})}]{PhysRevB.48.16373}%
  \BibitemOpen
  \bibfield  {author} {\bibinfo {author} {\bibfnamefont {R.~J.}\ \bibnamefont
  {Stoner}}\ and\ \bibinfo {author} {\bibfnamefont {H.~J.}\ \bibnamefont
  {Maris}},\ }\href {\doibase 10.1103/PhysRevB.48.16373} {\bibfield  {journal}
  {\bibinfo  {journal} {Phys. Rev. B}\ }\textbf {\bibinfo {volume} {48}},\
  \bibinfo {pages} {16373} (\bibinfo {year} {1993}{\natexlab{a}})}\BibitemShut
  {NoStop}%
\bibitem [{\citenamefont {Costescu}\ \emph {et~al.}(2003)\citenamefont
  {Costescu}, \citenamefont {Wall},\ and\ \citenamefont {Cahill}}]{RN6}%
  \BibitemOpen
  \bibfield  {author} {\bibinfo {author} {\bibfnamefont {R.~M.}\ \bibnamefont
  {Costescu}}, \bibinfo {author} {\bibfnamefont {M.~A.}\ \bibnamefont {Wall}},
  \ and\ \bibinfo {author} {\bibfnamefont {D.~G.}\ \bibnamefont {Cahill}},\
  }\href {\doibase 10.1103/PhysRevB.67.054302} {\bibfield  {journal} {\bibinfo
  {journal} {Physical Review B}\ }\textbf {\bibinfo {volume} {67}} (\bibinfo
  {year} {2003}),\ 10.1103/PhysRevB.67.054302}\BibitemShut {NoStop}%
\bibitem [{\citenamefont {Wilson}\ \emph {et~al.}(2015)\citenamefont {Wilson},
  \citenamefont {Apgar}, \citenamefont {Hsieh}, \citenamefont {Martin},\ and\
  \citenamefont {Cahill}}]{Wilson_PhysRevB.91.115414}%
  \BibitemOpen
  \bibfield  {author} {\bibinfo {author} {\bibfnamefont {R.~B.}\ \bibnamefont
  {Wilson}}, \bibinfo {author} {\bibfnamefont {B.~A.}\ \bibnamefont {Apgar}},
  \bibinfo {author} {\bibfnamefont {W.-P.}\ \bibnamefont {Hsieh}}, \bibinfo
  {author} {\bibfnamefont {L.~W.}\ \bibnamefont {Martin}}, \ and\ \bibinfo
  {author} {\bibfnamefont {D.~G.}\ \bibnamefont {Cahill}},\ }\href {\doibase
  10.1103/PhysRevB.91.115414} {\bibfield  {journal} {\bibinfo  {journal} {Phys.
  Rev. B}\ }\textbf {\bibinfo {volume} {91}},\ \bibinfo {pages} {115414}
  (\bibinfo {year} {2015})}\BibitemShut {NoStop}%
\bibitem [{\citenamefont {Minnich}\ \emph {et~al.}(2011)\citenamefont
  {Minnich}, \citenamefont {Johnson}, \citenamefont {Schmidt}, \citenamefont
  {Esfarjani}, \citenamefont {Dresselhaus}, \citenamefont {Nelson},\ and\
  \citenamefont {Chen}}]{Minnich_PhysRevLett.107.095901}%
  \BibitemOpen
  \bibfield  {author} {\bibinfo {author} {\bibfnamefont {A.~J.}\ \bibnamefont
  {Minnich}}, \bibinfo {author} {\bibfnamefont {J.~A.}\ \bibnamefont
  {Johnson}}, \bibinfo {author} {\bibfnamefont {A.~J.}\ \bibnamefont
  {Schmidt}}, \bibinfo {author} {\bibfnamefont {K.}~\bibnamefont {Esfarjani}},
  \bibinfo {author} {\bibfnamefont {M.~S.}\ \bibnamefont {Dresselhaus}},
  \bibinfo {author} {\bibfnamefont {K.~A.}\ \bibnamefont {Nelson}}, \ and\
  \bibinfo {author} {\bibfnamefont {G.}~\bibnamefont {Chen}},\ }\href {\doibase
  10.1103/PhysRevLett.107.095901} {\bibfield  {journal} {\bibinfo  {journal}
  {Phys. Rev. Lett.}\ }\textbf {\bibinfo {volume} {107}},\ \bibinfo {pages}
  {095901} (\bibinfo {year} {2011})}\BibitemShut {NoStop}%
\bibitem [{\citenamefont {Wilson}\ and\ \citenamefont {Cahill}(2014)}]{RN24}%
  \BibitemOpen
  \bibfield  {author} {\bibinfo {author} {\bibfnamefont {R.~B.}\ \bibnamefont
  {Wilson}}\ and\ \bibinfo {author} {\bibfnamefont {D.~G.}\ \bibnamefont
  {Cahill}},\ }\href {\doibase 10.1038/ncomms6075} {\bibfield  {journal}
  {\bibinfo  {journal} {Nat Commun}\ }\textbf {\bibinfo {volume} {5}},\
  \bibinfo {pages} {5075} (\bibinfo {year} {2014})}\BibitemShut {NoStop}%
\bibitem [{\citenamefont {Liu}\ \emph {et~al.}(2014)\citenamefont {Liu},
  \citenamefont {Xie}, \citenamefont {Yang}, \citenamefont {Li},\ and\
  \citenamefont {Thong}}]{RN7}%
  \BibitemOpen
  \bibfield  {author} {\bibinfo {author} {\bibfnamefont {D.}~\bibnamefont
  {Liu}}, \bibinfo {author} {\bibfnamefont {R.}~\bibnamefont {Xie}}, \bibinfo
  {author} {\bibfnamefont {N.}~\bibnamefont {Yang}}, \bibinfo {author}
  {\bibfnamefont {B.}~\bibnamefont {Li}}, \ and\ \bibinfo {author}
  {\bibfnamefont {J.~T.}\ \bibnamefont {Thong}},\ }\href {\doibase
  10.1021/nl4041516} {\bibfield  {journal} {\bibinfo  {journal} {Nano Lett}\
  }\textbf {\bibinfo {volume} {14}},\ \bibinfo {pages} {806} (\bibinfo {year}
  {2014})}\BibitemShut {NoStop}%
\bibitem [{\citenamefont {Ishiwara}\ \emph {et~al.}(1979)\citenamefont
  {Ishiwara}, \citenamefont {Hikosaka},\ and\ \citenamefont
  {Furukawa}}]{Ishiwara_JAP_1979_PtSi}%
  \BibitemOpen
  \bibfield  {author} {\bibinfo {author} {\bibfnamefont {H.}~\bibnamefont
  {Ishiwara}}, \bibinfo {author} {\bibfnamefont {K.}~\bibnamefont {Hikosaka}},
  \ and\ \bibinfo {author} {\bibfnamefont {S.}~\bibnamefont {Furukawa}},\
  }\href {\doibase http://dx.doi.org/10.1063/1.326628} {\bibfield  {journal}
  {\bibinfo  {journal} {Journal of Applied Physics}\ }\textbf {\bibinfo
  {volume} {50}},\ \bibinfo {pages} {5302} (\bibinfo {year}
  {1979})}\BibitemShut {NoStop}%
\bibitem [{\citenamefont {Dass}\ \emph {et~al.}(1991)\citenamefont {Dass},
  \citenamefont {Fraser},\ and\ \citenamefont {Wei}}]{Dass_APL_1991}%
  \BibitemOpen
  \bibfield  {author} {\bibinfo {author} {\bibfnamefont {M.~L.~A.}\
  \bibnamefont {Dass}}, \bibinfo {author} {\bibfnamefont {D.~B.}\ \bibnamefont
  {Fraser}}, \ and\ \bibinfo {author} {\bibfnamefont {C.-h.}\ \bibnamefont
  {Wei}},\ }\href {\doibase http://dx.doi.org/10.1063/1.104345} {\bibfield
  {journal} {\bibinfo  {journal} {Applied Physics Letters}\ }\textbf {\bibinfo
  {volume} {58}},\ \bibinfo {pages} {1308} (\bibinfo {year}
  {1991})}\BibitemShut {NoStop}%
\bibitem [{\citenamefont {D{'}Avitaya}\ \emph {et~al.}(1985)\citenamefont
  {D{'}Avitaya}, \citenamefont {Delage}, \citenamefont {Rosencher},\ and\
  \citenamefont {Derrien}}]{DAvitaya_JVac_1985}%
  \BibitemOpen
  \bibfield  {author} {\bibinfo {author} {\bibfnamefont {F.~A.}\ \bibnamefont
  {D{'}Avitaya}}, \bibinfo {author} {\bibfnamefont {S.}~\bibnamefont {Delage}},
  \bibinfo {author} {\bibfnamefont {E.}~\bibnamefont {Rosencher}}, \ and\
  \bibinfo {author} {\bibfnamefont {J.}~\bibnamefont {Derrien}},\ }\href
  {\doibase http://dx.doi.org/10.1116/1.583140} {\bibfield  {journal} {\bibinfo
   {journal} {Journal of Vacuum Science and Technology B}\ }\textbf {\bibinfo
  {volume} {3}},\ \bibinfo {pages} {770} (\bibinfo {year} {1985})}\BibitemShut
  {NoStop}%
\bibitem [{\citenamefont {Mangelinck}\ \emph {et~al.}(1999)\citenamefont
  {Mangelinck}, \citenamefont {Dai}, \citenamefont {Pan},\ and\ \citenamefont
  {Lahiri}}]{Mangelinck_APL_1999_75_12}%
  \BibitemOpen
  \bibfield  {author} {\bibinfo {author} {\bibfnamefont {D.}~\bibnamefont
  {Mangelinck}}, \bibinfo {author} {\bibfnamefont {J.~Y.}\ \bibnamefont {Dai}},
  \bibinfo {author} {\bibfnamefont {J.~S.}\ \bibnamefont {Pan}}, \ and\
  \bibinfo {author} {\bibfnamefont {S.~K.}\ \bibnamefont {Lahiri}},\ }\href
  {\doibase http://dx.doi.org/10.1063/1.124803} {\bibfield  {journal} {\bibinfo
   {journal} {Applied Physics Letters}\ }\textbf {\bibinfo {volume} {75}},\
  \bibinfo {pages} {1736} (\bibinfo {year} {1999})}\BibitemShut {NoStop}%
\bibitem [{\citenamefont {Wan}\ and\ \citenamefont {Wu}(1997)}]{WAN1997105}%
  \BibitemOpen
  \bibfield  {author} {\bibinfo {author} {\bibfnamefont {W.-K.}\ \bibnamefont
  {Wan}}\ and\ \bibinfo {author} {\bibfnamefont {S.-T.}\ \bibnamefont {Wu}},\
  }\href {\doibase http://dx.doi.org/10.1016/S0167-577X(96)00179-6} {\bibfield
  {journal} {\bibinfo  {journal} {Materials Letters}\ }\textbf {\bibinfo
  {volume} {30}},\ \bibinfo {pages} {105 } (\bibinfo {year}
  {1997})}\BibitemShut {NoStop}%
\bibitem [{\citenamefont {Yang}\ \emph {et~al.}(2003)\citenamefont {Yang},
  \citenamefont {Park}, \citenamefont {Park}, \citenamefont {Lim},
  \citenamefont {Lee}, \citenamefont {Park},\ and\ \citenamefont
  {Kim}}]{Yang_EpiC49_JAP_2003}%
  \BibitemOpen
  \bibfield  {author} {\bibinfo {author} {\bibfnamefont {J.-M.}\ \bibnamefont
  {Yang}}, \bibinfo {author} {\bibfnamefont {J.-C.}\ \bibnamefont {Park}},
  \bibinfo {author} {\bibfnamefont {D.-G.}\ \bibnamefont {Park}}, \bibinfo
  {author} {\bibfnamefont {K.-Y.}\ \bibnamefont {Lim}}, \bibinfo {author}
  {\bibfnamefont {S.-Y.}\ \bibnamefont {Lee}}, \bibinfo {author} {\bibfnamefont
  {S.-W.}\ \bibnamefont {Park}}, \ and\ \bibinfo {author} {\bibfnamefont
  {Y.-J.}\ \bibnamefont {Kim}},\ }\href {\doibase
  http://dx.doi.org/10.1063/1.1604955} {\bibfield  {journal} {\bibinfo
  {journal} {Journal of Applied Physics}\ }\textbf {\bibinfo {volume} {94}},\
  \bibinfo {pages} {4198} (\bibinfo {year} {2003})}\BibitemShut {NoStop}%
\bibitem [{\citenamefont {Chien}\ \emph {et~al.}(1985)\citenamefont {Chien},
  \citenamefont {Cheng}, \citenamefont {Nieh},\ and\ \citenamefont
  {Chen}}]{:/content/aip/journal/jap/57/6/10.1063/1.334420}%
  \BibitemOpen
  \bibfield  {author} {\bibinfo {author} {\bibfnamefont {C.~J.}\ \bibnamefont
  {Chien}}, \bibinfo {author} {\bibfnamefont {H.~C.}\ \bibnamefont {Cheng}},
  \bibinfo {author} {\bibfnamefont {C.~W.}\ \bibnamefont {Nieh}}, \ and\
  \bibinfo {author} {\bibfnamefont {L.~J.}\ \bibnamefont {Chen}},\ }\href
  {\doibase http://dx.doi.org/10.1063/1.334420} {\bibfield  {journal} {\bibinfo
   {journal} {Journal of Applied Physics}\ }\textbf {\bibinfo {volume} {57}},\
  \bibinfo {pages} {1887} (\bibinfo {year} {1985})}\BibitemShut {NoStop}%
\bibitem [{\citenamefont {Kim}\ \emph {et~al.}(1999)\citenamefont {Kim},
  \citenamefont {Kang}, \citenamefont {Choi}, \citenamefont {Lee},\ and\
  \citenamefont {Olson}}]{Kim19998}%
  \BibitemOpen
  \bibfield  {author} {\bibinfo {author} {\bibfnamefont {K.}~\bibnamefont
  {Kim}}, \bibinfo {author} {\bibfnamefont {J.-S.}\ \bibnamefont {Kang}},
  \bibinfo {author} {\bibfnamefont {C.}~\bibnamefont {Choi}}, \bibinfo {author}
  {\bibfnamefont {J.}~\bibnamefont {Lee}}, \ and\ \bibinfo {author}
  {\bibfnamefont {C.}~\bibnamefont {Olson}},\ }\href {\doibase
  http://dx.doi.org/10.1016/S0169-4332(99)00251-2} {\bibfield  {journal}
  {\bibinfo  {journal} {Applied Surface Science}\ }\textbf {\bibinfo {volume}
  {150}},\ \bibinfo {pages} {8 } (\bibinfo {year} {1999})}\BibitemShut
  {NoStop}%
\bibitem [{\citenamefont {Grimaldi}\ \emph {et~al.}(1994)\citenamefont
  {Grimaldi}, \citenamefont {FranzÃ²}, \citenamefont {Ravesi}, \citenamefont
  {Terrasi}, \citenamefont {Spinella},\ and\ \citenamefont
  {Mantia}}]{GRIMALDI199419}%
  \BibitemOpen
  \bibfield  {author} {\bibinfo {author} {\bibfnamefont {M.}~\bibnamefont
  {Grimaldi}}, \bibinfo {author} {\bibfnamefont {G.}~\bibnamefont {FranzÃ²}},
  \bibinfo {author} {\bibfnamefont {S.}~\bibnamefont {Ravesi}}, \bibinfo
  {author} {\bibfnamefont {A.}~\bibnamefont {Terrasi}}, \bibinfo {author}
  {\bibfnamefont {C.}~\bibnamefont {Spinella}}, \ and\ \bibinfo {author}
  {\bibfnamefont {A.~L.}\ \bibnamefont {Mantia}},\ }\href {\doibase
  http://dx.doi.org/10.1016/0169-4332(94)90095-7} {\bibfield  {journal}
  {\bibinfo  {journal} {Applied Surface Science}\ }\textbf {\bibinfo {volume}
  {74}},\ \bibinfo {pages} {19 } (\bibinfo {year} {1994})}\BibitemShut
  {NoStop}%
\bibitem [{\citenamefont {Gurvitch}\ \emph {et~al.}(1987)\citenamefont
  {Gurvitch}, \citenamefont {Levi}, \citenamefont {Tung},\ and\ \citenamefont
  {Nakahara}}]{Gurvitch_APL_1987}%
  \BibitemOpen
  \bibfield  {author} {\bibinfo {author} {\bibfnamefont {M.}~\bibnamefont
  {Gurvitch}}, \bibinfo {author} {\bibfnamefont {A.~F.~J.}\ \bibnamefont
  {Levi}}, \bibinfo {author} {\bibfnamefont {R.~T.}\ \bibnamefont {Tung}}, \
  and\ \bibinfo {author} {\bibfnamefont {S.}~\bibnamefont {Nakahara}},\ }\href
  {\doibase http://dx.doi.org/10.1063/1.98453} {\bibfield  {journal} {\bibinfo
  {journal} {Applied Physics Letters}\ }\textbf {\bibinfo {volume} {51}},\
  \bibinfo {pages} {311} (\bibinfo {year} {1987})}\BibitemShut {NoStop}%
\bibitem [{\citenamefont {Knapp}\ and\ \citenamefont
  {Picraux}(1986)}]{Knapp_APL_1986_REsilicides}%
  \BibitemOpen
  \bibfield  {author} {\bibinfo {author} {\bibfnamefont {J.~A.}\ \bibnamefont
  {Knapp}}\ and\ \bibinfo {author} {\bibfnamefont {S.~T.}\ \bibnamefont
  {Picraux}},\ }\href {\doibase http://dx.doi.org/10.1063/1.96532} {\bibfield
  {journal} {\bibinfo  {journal} {Applied Physics Letters}\ }\textbf {\bibinfo
  {volume} {48}},\ \bibinfo {pages} {466} (\bibinfo {year} {1986})}\BibitemShut
  {NoStop}%
\bibitem [{\citenamefont {Lin}\ and\ \citenamefont
  {Chen}(1985)}]{Lin_APL_epiMoSi2}%
  \BibitemOpen
  \bibfield  {author} {\bibinfo {author} {\bibfnamefont {W.~T.}\ \bibnamefont
  {Lin}}\ and\ \bibinfo {author} {\bibfnamefont {L.~J.}\ \bibnamefont {Chen}},\
  }\href {\doibase http://dx.doi.org/10.1063/1.95760} {\bibfield  {journal}
  {\bibinfo  {journal} {Applied Physics Letters}\ }\textbf {\bibinfo {volume}
  {46}},\ \bibinfo {pages} {1061} (\bibinfo {year} {1985})}\BibitemShut
  {NoStop}%
\bibitem [{\citenamefont {Kircher}(1971)}]{KIRCHER1971507}%
  \BibitemOpen
  \bibfield  {author} {\bibinfo {author} {\bibfnamefont {C.}~\bibnamefont
  {Kircher}},\ }\href {\doibase http://dx.doi.org/10.1016/0038-1101(71)90061-X}
  {\bibfield  {journal} {\bibinfo  {journal} {Solid-State Electronics}\
  }\textbf {\bibinfo {volume} {14}},\ \bibinfo {pages} {507 } (\bibinfo {year}
  {1971})}\BibitemShut {NoStop}%
\bibitem [{\citenamefont {Wu}\ \emph {et~al.}(1987)\citenamefont {Wu},
  \citenamefont {Chu},\ and\ \citenamefont {Chen}}]{Wu_JAP_1987_TaSi2}%
  \BibitemOpen
  \bibfield  {author} {\bibinfo {author} {\bibfnamefont {I.~C.}\ \bibnamefont
  {Wu}}, \bibinfo {author} {\bibfnamefont {J.~J.}\ \bibnamefont {Chu}}, \ and\
  \bibinfo {author} {\bibfnamefont {L.~J.}\ \bibnamefont {Chen}},\ }\href
  {\doibase http://dx.doi.org/10.1063/1.339693} {\bibfield  {journal} {\bibinfo
   {journal} {Journal of Applied Physics}\ }\textbf {\bibinfo {volume} {62}},\
  \bibinfo {pages} {879} (\bibinfo {year} {1987})}\BibitemShut {NoStop}%
\bibitem [{\citenamefont {Lin}\ and\ \citenamefont
  {Chen}(1986)}]{Lin_JAP_1986_WSi2}%
  \BibitemOpen
  \bibfield  {author} {\bibinfo {author} {\bibfnamefont {W.~T.}\ \bibnamefont
  {Lin}}\ and\ \bibinfo {author} {\bibfnamefont {L.~J.}\ \bibnamefont {Chen}},\
  }\href {\doibase http://dx.doi.org/10.1063/1.336818} {\bibfield  {journal}
  {\bibinfo  {journal} {Journal of Applied Physics}\ }\textbf {\bibinfo
  {volume} {59}},\ \bibinfo {pages} {3481} (\bibinfo {year}
  {1986})}\BibitemShut {NoStop}%
\bibitem [{\citenamefont {Chang}\ and\ \citenamefont
  {Chou}(1989)}]{Chang_JAP_1989_OsSi2}%
  \BibitemOpen
  \bibfield  {author} {\bibinfo {author} {\bibfnamefont {Y.~S.}\ \bibnamefont
  {Chang}}\ and\ \bibinfo {author} {\bibfnamefont {M.~L.}\ \bibnamefont
  {Chou}},\ }\href {\doibase http://dx.doi.org/10.1063/1.344185} {\bibfield
  {journal} {\bibinfo  {journal} {Journal of Applied Physics}\ }\textbf
  {\bibinfo {volume} {66}},\ \bibinfo {pages} {3011} (\bibinfo {year}
  {1989})}\BibitemShut {NoStop}%
\bibitem [{\citenamefont {H\"ormann}\ \emph {et~al.}(2009)\citenamefont
  {H\"ormann}, \citenamefont {Remmele}, \citenamefont {Klepeis}, \citenamefont
  {Pankratov}, \citenamefont {Gr\"unleitner}, \citenamefont {Schulz},
  \citenamefont {Falke},\ and\ \citenamefont {Bleloch}}]{PhysRevB.79.104116}%
  \BibitemOpen
  \bibfield  {author} {\bibinfo {author} {\bibfnamefont {U.}~\bibnamefont
  {H\"ormann}}, \bibinfo {author} {\bibfnamefont {T.}~\bibnamefont {Remmele}},
  \bibinfo {author} {\bibfnamefont {J.~E.}\ \bibnamefont {Klepeis}}, \bibinfo
  {author} {\bibfnamefont {O.}~\bibnamefont {Pankratov}}, \bibinfo {author}
  {\bibfnamefont {H.}~\bibnamefont {Gr\"unleitner}}, \bibinfo {author}
  {\bibfnamefont {M.}~\bibnamefont {Schulz}}, \bibinfo {author} {\bibfnamefont
  {M.}~\bibnamefont {Falke}}, \ and\ \bibinfo {author} {\bibfnamefont
  {A.}~\bibnamefont {Bleloch}},\ }\href {\doibase 10.1103/PhysRevB.79.104116}
  {\bibfield  {journal} {\bibinfo  {journal} {Phys. Rev. B}\ }\textbf {\bibinfo
  {volume} {79}},\ \bibinfo {pages} {104116} (\bibinfo {year}
  {2009})}\BibitemShut {NoStop}%
\bibitem [{\citenamefont {Fathauer}\ \emph {et~al.}(1990)\citenamefont
  {Fathauer}, \citenamefont {Xiao}, \citenamefont {Hashimoto},\ and\
  \citenamefont {Nieh}}]{RN8}%
  \BibitemOpen
  \bibfield  {author} {\bibinfo {author} {\bibfnamefont {R.~W.}\ \bibnamefont
  {Fathauer}}, \bibinfo {author} {\bibfnamefont {Q.~F.}\ \bibnamefont {Xiao}},
  \bibinfo {author} {\bibfnamefont {S.}~\bibnamefont {Hashimoto}}, \ and\
  \bibinfo {author} {\bibfnamefont {C.~W.}\ \bibnamefont {Nieh}},\ }\href
  {\doibase 10.1063/1.103592} {\bibfield  {journal} {\bibinfo  {journal}
  {Applied Physics Letters}\ }\textbf {\bibinfo {volume} {57}},\ \bibinfo
  {pages} {686} (\bibinfo {year} {1990})}\BibitemShut {NoStop}%
\bibitem [{\citenamefont {Kawarada}\ \emph {et~al.}(1986)\citenamefont
  {Kawarada}, \citenamefont {Ishida}, \citenamefont {Nakanishi}, \citenamefont
  {I.},\ and\ \citenamefont {S.}}]{RN9}%
  \BibitemOpen
  \bibfield  {author} {\bibinfo {author} {\bibfnamefont {H.}~\bibnamefont
  {Kawarada}}, \bibinfo {author} {\bibfnamefont {M.}~\bibnamefont {Ishida}},
  \bibinfo {author} {\bibfnamefont {J.}~\bibnamefont {Nakanishi}}, \bibinfo
  {author} {\bibfnamefont {O.}~\bibnamefont {I.}}, \ and\ \bibinfo {author}
  {\bibfnamefont {H.}~\bibnamefont {S.}},\ }\href@noop {} {\bibfield  {journal}
  {\bibinfo  {journal} {Philosophical Magazine A}\ }\textbf {\bibinfo {volume}
  {54}},\ \bibinfo {pages} {729} (\bibinfo {year} {1986})}\BibitemShut
  {NoStop}%
\bibitem [{\citenamefont {Kawarada}\ \emph
  {et~al.}(1984{\natexlab{a}})\citenamefont {Kawarada}, \citenamefont
  {Ohdomari},\ and\ \citenamefont {Horiuchi}}]{RN10}%
  \BibitemOpen
  \bibfield  {author} {\bibinfo {author} {\bibfnamefont {H.}~\bibnamefont
  {Kawarada}}, \bibinfo {author} {\bibfnamefont {I.}~\bibnamefont {Ohdomari}},
  \ and\ \bibinfo {author} {\bibfnamefont {S.}~\bibnamefont {Horiuchi}},\
  }\href@noop {} {\bibfield  {journal} {\bibinfo  {journal} {Japanese Journal
  of Applied Physics}\ }\textbf {\bibinfo {volume} {23}},\ \bibinfo {pages}
  {L799} (\bibinfo {year} {1984}{\natexlab{a}})}\BibitemShut {NoStop}%
\bibitem [{\citenamefont {Wawro}\ \emph {et~al.}(2005)\citenamefont {Wawro},
  \citenamefont {Suto},\ and\ \citenamefont {Kasuya}}]{RN11}%
  \BibitemOpen
  \bibfield  {author} {\bibinfo {author} {\bibfnamefont {A.}~\bibnamefont
  {Wawro}}, \bibinfo {author} {\bibfnamefont {S.}~\bibnamefont {Suto}}, \ and\
  \bibinfo {author} {\bibfnamefont {A.}~\bibnamefont {Kasuya}},\ }\href
  {\doibase 10.1103/PhysRevB.72.205302} {\bibfield  {journal} {\bibinfo
  {journal} {Physical Review B}\ }\textbf {\bibinfo {volume} {72}},\ \bibinfo
  {pages} {205302} (\bibinfo {year} {2005})}\BibitemShut {NoStop}%
\bibitem [{\citenamefont {Liehr}\ \emph {et~al.}(1985)\citenamefont {Liehr},
  \citenamefont {Schmid}, \citenamefont {LeGoues},\ and\ \citenamefont
  {Ho}}]{RN12}%
  \BibitemOpen
  \bibfield  {author} {\bibinfo {author} {\bibfnamefont {M.}~\bibnamefont
  {Liehr}}, \bibinfo {author} {\bibfnamefont {P.~E.}\ \bibnamefont {Schmid}},
  \bibinfo {author} {\bibfnamefont {F.~K.}\ \bibnamefont {LeGoues}}, \ and\
  \bibinfo {author} {\bibfnamefont {P.~S.}\ \bibnamefont {Ho}},\ }\href
  {\doibase 10.1103/PhysRevLett.54.2139} {\bibfield  {journal} {\bibinfo
  {journal} {Physical Review Letters}\ }\textbf {\bibinfo {volume} {54}},\
  \bibinfo {pages} {2139} (\bibinfo {year} {1985})}\BibitemShut {NoStop}%
\bibitem [{\citenamefont {d{'}Heurle}\ \emph {et~al.}(1984)\citenamefont
  {d{'}Heurle}, \citenamefont {Petersson}, \citenamefont {Baglin},
  \citenamefont {La~Placa},\ and\ \citenamefont {Wong}}]{RN13}%
  \BibitemOpen
  \bibfield  {author} {\bibinfo {author} {\bibfnamefont {F.}~\bibnamefont
  {d{'}Heurle}}, \bibinfo {author} {\bibfnamefont {C.~S.}\ \bibnamefont
  {Petersson}}, \bibinfo {author} {\bibfnamefont {J.~E.~E.}\ \bibnamefont
  {Baglin}}, \bibinfo {author} {\bibfnamefont {S.~J.}\ \bibnamefont
  {La~Placa}}, \ and\ \bibinfo {author} {\bibfnamefont {C.~Y.}\ \bibnamefont
  {Wong}},\ }\href {\doibase 10.1063/1.333021} {\bibfield  {journal} {\bibinfo
  {journal} {Journal of Applied Physics}\ }\textbf {\bibinfo {volume} {55}},\
  \bibinfo {pages} {4208} (\bibinfo {year} {1984})}\BibitemShut {NoStop}%
\bibitem [{\citenamefont {F{\"o}ll}\ \emph {et~al.}(1982)\citenamefont
  {F{\"o}ll}, \citenamefont {Ho},\ and\ \citenamefont {Tu}}]{RN14}%
  \BibitemOpen
  \bibfield  {author} {\bibinfo {author} {\bibfnamefont {H.}~\bibnamefont
  {F{\"o}ll}}, \bibinfo {author} {\bibfnamefont {P.~S.}\ \bibnamefont {Ho}}, \
  and\ \bibinfo {author} {\bibfnamefont {K.~N.}\ \bibnamefont {Tu}},\
  }\href@noop {} {\bibfield  {journal} {\bibinfo  {journal} {Philosophical
  Magazine A}\ }\textbf {\bibinfo {volume} {45}},\ \bibinfo {pages} {31}
  (\bibinfo {year} {1982})}\BibitemShut {NoStop}%
\bibitem [{\citenamefont {Kawarada}\ \emph
  {et~al.}(1984{\natexlab{b}})\citenamefont {Kawarada}, \citenamefont
  {Ohdomari},\ and\ \citenamefont {Horiuchi}}]{Kawarada}%
  \BibitemOpen
  \bibfield  {author} {\bibinfo {author} {\bibfnamefont {H.}~\bibnamefont
  {Kawarada}}, \bibinfo {author} {\bibfnamefont {I.}~\bibnamefont {Ohdomari}},
  \ and\ \bibinfo {author} {\bibfnamefont {S.}~\bibnamefont {Horiuchi}},\
  }\href {http://stacks.iop.org/1347-4065/23/i=10A/a=L799} {\bibfield
  {journal} {\bibinfo  {journal} {Japanese Journal of Applied Physics}\
  }\textbf {\bibinfo {volume} {23}},\ \bibinfo {pages} {L799} (\bibinfo {year}
  {1984}{\natexlab{b}})}\BibitemShut {NoStop}%
\bibitem [{\citenamefont {Kang}\ \emph {et~al.}(2008)\citenamefont {Kang},
  \citenamefont {Koh}, \citenamefont {Chiritescu}, \citenamefont {Zheng},\ and\
  \citenamefont {Cahill}}]{Kang_Cahill_twotint_RSI_2008}%
  \BibitemOpen
  \bibfield  {author} {\bibinfo {author} {\bibfnamefont {K.}~\bibnamefont
  {Kang}}, \bibinfo {author} {\bibfnamefont {Y.~K.}\ \bibnamefont {Koh}},
  \bibinfo {author} {\bibfnamefont {C.}~\bibnamefont {Chiritescu}}, \bibinfo
  {author} {\bibfnamefont {X.}~\bibnamefont {Zheng}}, \ and\ \bibinfo {author}
  {\bibfnamefont {D.~G.}\ \bibnamefont {Cahill}},\ }\href {\doibase
  http://dx.doi.org/10.1063/1.3020759} {\bibfield  {journal} {\bibinfo
  {journal} {Review of Scientific Instruments}\ }\textbf {\bibinfo {volume}
  {79}},\ \bibinfo {pages} {114901} (\bibinfo {year} {2008})}\BibitemShut
  {NoStop}%
\bibitem [{\citenamefont {Cahill}(2004)}]{Cahill_TDTR_RSI_2004}%
  \BibitemOpen
  \bibfield  {author} {\bibinfo {author} {\bibfnamefont {D.~G.}\ \bibnamefont
  {Cahill}},\ }\href {\doibase http://dx.doi.org/10.1063/1.1819431} {\bibfield
  {journal} {\bibinfo  {journal} {Review of Scientific Instruments}\ }\textbf
  {\bibinfo {volume} {75}},\ \bibinfo {pages} {5119} (\bibinfo {year}
  {2004})}\BibitemShut {NoStop}%
\bibitem [{\citenamefont {Ditchek}(1984)}]{CoSi2_temp}%
  \BibitemOpen
  \bibfield  {author} {\bibinfo {author} {\bibfnamefont {B.~M.}\ \bibnamefont
  {Ditchek}},\ }\href {\doibase http://dx.doi.org/10.1016/0022-0248(84)90031-9}
  {\bibfield  {journal} {\bibinfo  {journal} {Journal of Crystal Growth}\
  }\textbf {\bibinfo {volume} {69}},\ \bibinfo {pages} {207} (\bibinfo {year}
  {1984})}\BibitemShut {NoStop}%
\bibitem [{\citenamefont {Malhotra}\ \emph {et~al.}(1984)\citenamefont
  {Malhotra}, \citenamefont {Martin},\ and\ \citenamefont
  {Mahan}}]{TiSi2_temp}%
  \BibitemOpen
  \bibfield  {author} {\bibinfo {author} {\bibfnamefont {V.}~\bibnamefont
  {Malhotra}}, \bibinfo {author} {\bibfnamefont {T.~L.}\ \bibnamefont
  {Martin}}, \ and\ \bibinfo {author} {\bibfnamefont {J.~E.}\ \bibnamefont
  {Mahan}},\ }\href {\doibase http://dx.doi.org/10.1116/1.582905} {\bibfield
  {journal} {\bibinfo  {journal} {Journal of Vacuum Science \& Technology B}\
  }\textbf {\bibinfo {volume} {2}},\ \bibinfo {pages} {10} (\bibinfo {year}
  {1984})}\BibitemShut {NoStop}%
\bibitem [{\citenamefont {Meyer}\ \emph {et~al.}(1997)\citenamefont {Meyer},
  \citenamefont {Gottlieb}, \citenamefont {Laborde}, \citenamefont {Yang},
  \citenamefont {Lasjaunias}, \citenamefont {Sulpice},\ and\ \citenamefont
  {Madar}}]{NiSi_temp}%
  \BibitemOpen
  \bibfield  {author} {\bibinfo {author} {\bibfnamefont {B.}~\bibnamefont
  {Meyer}}, \bibinfo {author} {\bibfnamefont {U.}~\bibnamefont {Gottlieb}},
  \bibinfo {author} {\bibfnamefont {O.}~\bibnamefont {Laborde}}, \bibinfo
  {author} {\bibfnamefont {H.}~\bibnamefont {Yang}}, \bibinfo {author}
  {\bibfnamefont {J.}~\bibnamefont {Lasjaunias}}, \bibinfo {author}
  {\bibfnamefont {A.}~\bibnamefont {Sulpice}}, \ and\ \bibinfo {author}
  {\bibfnamefont {R.}~\bibnamefont {Madar}},\ }\href {\doibase
  http://dx.doi.org.udel.idm.oclc.org/10.1016/S0167-9317(97)00155-X} {\bibfield
   {journal} {\bibinfo  {journal} {Microelectronic Engineering}\ }\textbf
  {\bibinfo {volume} {37-38}},\ \bibinfo {pages} {523} (\bibinfo {year}
  {1997})}\BibitemShut {NoStop}%
\bibitem [{\citenamefont {Wang}\ \emph {et~al.}(2010)\citenamefont {Wang},
  \citenamefont {Park}, \citenamefont {Koh},\ and\ \citenamefont
  {Cahill}}]{YuxinWang_thermoreflectance_coeff_2010}%
  \BibitemOpen
  \bibfield  {author} {\bibinfo {author} {\bibfnamefont {Y.}~\bibnamefont
  {Wang}}, \bibinfo {author} {\bibfnamefont {J.~Y.}\ \bibnamefont {Park}},
  \bibinfo {author} {\bibfnamefont {Y.~K.}\ \bibnamefont {Koh}}, \ and\
  \bibinfo {author} {\bibfnamefont {D.~G.}\ \bibnamefont {Cahill}},\ }\href
  {\doibase http://dx.doi.org/10.1063/1.3457151} {\bibfield  {journal}
  {\bibinfo  {journal} {Journal of Applied Physics}\ }\textbf {\bibinfo
  {volume} {108}},\ \bibinfo {pages} {043507} (\bibinfo {year}
  {2010})}\BibitemShut {NoStop}%
\bibitem [{\citenamefont {Reddy}\ \emph {et~al.}(2005)\citenamefont {Reddy},
  \citenamefont {Castelino},\ and\ \citenamefont {Majumdar}}]{RN18}%
  \BibitemOpen
  \bibfield  {author} {\bibinfo {author} {\bibfnamefont {P.}~\bibnamefont
  {Reddy}}, \bibinfo {author} {\bibfnamefont {K.}~\bibnamefont {Castelino}}, \
  and\ \bibinfo {author} {\bibfnamefont {A.}~\bibnamefont {Majumdar}},\ }\href
  {\doibase 10.1063/1.2133890} {\bibfield  {journal} {\bibinfo  {journal}
  {Applied Physics Letters}\ }\textbf {\bibinfo {volume} {87}},\ \bibinfo
  {pages} {211908} (\bibinfo {year} {2005})}\BibitemShut {NoStop}%
\bibitem [{\citenamefont {Giannozzi}\ \emph {et~al.}(2009)\citenamefont
  {Giannozzi}, \citenamefont {Baroni}, \citenamefont {Bonini}, \citenamefont
  {Calandra}, \citenamefont {Car}, \citenamefont {Cavazzoni}, \citenamefont
  {Ceresoli}, \citenamefont {Chiarotti}, \citenamefont {Cococcioni},
  \citenamefont {Dabo}, \citenamefont {Dal~Corso}, \citenamefont
  {de~Gironcoli}, \citenamefont {Fabris}, \citenamefont {Fratesi},
  \citenamefont {Gebauer}, \citenamefont {Gerstmann}, \citenamefont
  {Gougoussis}, \citenamefont {Kokalj}, \citenamefont {Lazzeri}, \citenamefont
  {Martin-Samos}, \citenamefont {Marzari}, \citenamefont {Mauri}, \citenamefont
  {Mazzarello}, \citenamefont {Paolini}, \citenamefont {Pasquarello},
  \citenamefont {Paulatto}, \citenamefont {Sbraccia}, \citenamefont {Scandolo},
  \citenamefont {Sclauzero}, \citenamefont {Seitsonen}, \citenamefont
  {Smogunov}, \citenamefont {Umari},\ and\ \citenamefont
  {Wentzcovitch}}]{RN25}%
  \BibitemOpen
  \bibfield  {author} {\bibinfo {author} {\bibfnamefont {P.}~\bibnamefont
  {Giannozzi}}, \bibinfo {author} {\bibfnamefont {S.}~\bibnamefont {Baroni}},
  \bibinfo {author} {\bibfnamefont {N.}~\bibnamefont {Bonini}}, \bibinfo
  {author} {\bibfnamefont {M.}~\bibnamefont {Calandra}}, \bibinfo {author}
  {\bibfnamefont {R.}~\bibnamefont {Car}}, \bibinfo {author} {\bibfnamefont
  {C.}~\bibnamefont {Cavazzoni}}, \bibinfo {author} {\bibfnamefont
  {D.}~\bibnamefont {Ceresoli}}, \bibinfo {author} {\bibfnamefont {G.~L.}\
  \bibnamefont {Chiarotti}}, \bibinfo {author} {\bibfnamefont {M.}~\bibnamefont
  {Cococcioni}}, \bibinfo {author} {\bibfnamefont {I.}~\bibnamefont {Dabo}},
  \bibinfo {author} {\bibfnamefont {A.}~\bibnamefont {Dal~Corso}}, \bibinfo
  {author} {\bibfnamefont {S.}~\bibnamefont {de~Gironcoli}}, \bibinfo {author}
  {\bibfnamefont {S.}~\bibnamefont {Fabris}}, \bibinfo {author} {\bibfnamefont
  {G.}~\bibnamefont {Fratesi}}, \bibinfo {author} {\bibfnamefont
  {R.}~\bibnamefont {Gebauer}}, \bibinfo {author} {\bibfnamefont
  {U.}~\bibnamefont {Gerstmann}}, \bibinfo {author} {\bibfnamefont
  {C.}~\bibnamefont {Gougoussis}}, \bibinfo {author} {\bibfnamefont
  {A.}~\bibnamefont {Kokalj}}, \bibinfo {author} {\bibfnamefont
  {M.}~\bibnamefont {Lazzeri}}, \bibinfo {author} {\bibfnamefont
  {L.}~\bibnamefont {Martin-Samos}}, \bibinfo {author} {\bibfnamefont
  {N.}~\bibnamefont {Marzari}}, \bibinfo {author} {\bibfnamefont
  {F.}~\bibnamefont {Mauri}}, \bibinfo {author} {\bibfnamefont
  {R.}~\bibnamefont {Mazzarello}}, \bibinfo {author} {\bibfnamefont
  {S.}~\bibnamefont {Paolini}}, \bibinfo {author} {\bibfnamefont
  {A.}~\bibnamefont {Pasquarello}}, \bibinfo {author} {\bibfnamefont
  {L.}~\bibnamefont {Paulatto}}, \bibinfo {author} {\bibfnamefont
  {C.}~\bibnamefont {Sbraccia}}, \bibinfo {author} {\bibfnamefont
  {S.}~\bibnamefont {Scandolo}}, \bibinfo {author} {\bibfnamefont
  {G.}~\bibnamefont {Sclauzero}}, \bibinfo {author} {\bibfnamefont {A.~P.}\
  \bibnamefont {Seitsonen}}, \bibinfo {author} {\bibfnamefont {A.}~\bibnamefont
  {Smogunov}}, \bibinfo {author} {\bibfnamefont {P.}~\bibnamefont {Umari}}, \
  and\ \bibinfo {author} {\bibfnamefont {R.~M.}\ \bibnamefont {Wentzcovitch}},\
  }\href {\doibase 10.1088/0953-8984/21/39/395502} {\bibfield  {journal}
  {\bibinfo  {journal} {J Phys Condens Matter}\ }\textbf {\bibinfo {volume}
  {21}},\ \bibinfo {pages} {395502} (\bibinfo {year} {2009})}\BibitemShut
  {NoStop}%
\bibitem [{\citenamefont {Wardle}\ \emph {et~al.}(2005)\citenamefont {Wardle},
  \citenamefont {Goss}, \citenamefont {Briddon},\ and\ \citenamefont
  {Jones}}]{RN26}%
  \BibitemOpen
  \bibfield  {author} {\bibinfo {author} {\bibfnamefont {M.~G.}\ \bibnamefont
  {Wardle}}, \bibinfo {author} {\bibfnamefont {J.~P.}\ \bibnamefont {Goss}},
  \bibinfo {author} {\bibfnamefont {P.~R.}\ \bibnamefont {Briddon}}, \ and\
  \bibinfo {author} {\bibfnamefont {R.}~\bibnamefont {Jones}},\ }\href
  {\doibase 10.1002/pssa.200460508} {\bibfield  {journal} {\bibinfo  {journal}
  {physica status solidi (a)}\ }\textbf {\bibinfo {volume} {202}},\ \bibinfo
  {pages} {883} (\bibinfo {year} {2005})}\BibitemShut {NoStop}%
\bibitem [{\citenamefont {Stoner}\ and\ \citenamefont
  {Maris}(1993{\natexlab{b}})}]{RN19}%
  \BibitemOpen
  \bibfield  {author} {\bibinfo {author} {\bibfnamefont {R.~J.}\ \bibnamefont
  {Stoner}}\ and\ \bibinfo {author} {\bibfnamefont {H.~J.}\ \bibnamefont
  {Maris}},\ }\href {\doibase 10.1103/PhysRevB.48.16373} {\bibfield  {journal}
  {\bibinfo  {journal} {Physical Review B}\ }\textbf {\bibinfo {volume} {48}},\
  \bibinfo {pages} {16373} (\bibinfo {year} {1993}{\natexlab{b}})}\BibitemShut
  {NoStop}%
\bibitem [{\citenamefont {Perdew}\ \emph {et~al.}(2008)\citenamefont {Perdew},
  \citenamefont {Ruzsinszky}, \citenamefont {Csonka}, \citenamefont {Vydrov},
  \citenamefont {Scuseria}, \citenamefont {Constantin}, \citenamefont {Zhou},\
  and\ \citenamefont {Burke}}]{PBEsol}%
  \BibitemOpen
  \bibfield  {author} {\bibinfo {author} {\bibfnamefont {J.~P.}\ \bibnamefont
  {Perdew}}, \bibinfo {author} {\bibfnamefont {A.}~\bibnamefont {Ruzsinszky}},
  \bibinfo {author} {\bibfnamefont {G.~I.}\ \bibnamefont {Csonka}}, \bibinfo
  {author} {\bibfnamefont {O.~A.}\ \bibnamefont {Vydrov}}, \bibinfo {author}
  {\bibfnamefont {G.~E.}\ \bibnamefont {Scuseria}}, \bibinfo {author}
  {\bibfnamefont {L.~A.}\ \bibnamefont {Constantin}}, \bibinfo {author}
  {\bibfnamefont {X.}~\bibnamefont {Zhou}}, \ and\ \bibinfo {author}
  {\bibfnamefont {K.}~\bibnamefont {Burke}},\ }\href {\doibase
  10.1103/PhysRevLett.100.136406} {\bibfield  {journal} {\bibinfo  {journal}
  {Phys. Rev. Lett.}\ }\textbf {\bibinfo {volume} {100}},\ \bibinfo {pages}
  {136406} (\bibinfo {year} {2008})}\BibitemShut {NoStop}%
\bibitem [{\citenamefont {Kresse}\ and\ \citenamefont {Hafner}(1993)}]{VASP-1}%
  \BibitemOpen
  \bibfield  {author} {\bibinfo {author} {\bibfnamefont {G.}~\bibnamefont
  {Kresse}}\ and\ \bibinfo {author} {\bibfnamefont {J.}~\bibnamefont
  {Hafner}},\ }\href {\doibase 10.1103/PhysRevB.47.558} {\bibfield  {journal}
  {\bibinfo  {journal} {Phys. Rev. B}\ }\textbf {\bibinfo {volume} {47}},\
  \bibinfo {pages} {558} (\bibinfo {year} {1993})}\BibitemShut {NoStop}%
\bibitem [{\citenamefont {Kresse}\ and\ \citenamefont
  {Furthmüller}(1996)}]{VASP-2}%
  \BibitemOpen
  \bibfield  {author} {\bibinfo {author} {\bibfnamefont {G.}~\bibnamefont
  {Kresse}}\ and\ \bibinfo {author} {\bibfnamefont {J.}~\bibnamefont
  {Furthmüller}},\ }\href {\doibase
  http://dx.doi.org/10.1016/0927-0256(96)00008-0} {\bibfield  {journal}
  {\bibinfo  {journal} {Computational Materials Science}\ }\textbf {\bibinfo
  {volume} {6}},\ \bibinfo {pages} {15 } (\bibinfo {year} {1996})}\BibitemShut
  {NoStop}%
\bibitem [{\citenamefont {Kresse}\ and\ \citenamefont {Joubert}(1999)}]{PAW}%
  \BibitemOpen
  \bibfield  {author} {\bibinfo {author} {\bibfnamefont {G.}~\bibnamefont
  {Kresse}}\ and\ \bibinfo {author} {\bibfnamefont {D.}~\bibnamefont
  {Joubert}},\ }\href {\doibase 10.1103/PhysRevB.59.1758} {\bibfield  {journal}
  {\bibinfo  {journal} {Phys. Rev. B}\ }\textbf {\bibinfo {volume} {59}},\
  \bibinfo {pages} {1758} (\bibinfo {year} {1999})}\BibitemShut {NoStop}%
\bibitem [{\citenamefont {Graeber}\ \emph {et~al.}(1973)\citenamefont
  {Graeber}, \citenamefont {Baughman},\ and\ \citenamefont {Morosin}}]{PtSi}%
  \BibitemOpen
  \bibfield  {author} {\bibinfo {author} {\bibfnamefont {E.~J.}\ \bibnamefont
  {Graeber}}, \bibinfo {author} {\bibfnamefont {R.~J.}\ \bibnamefont
  {Baughman}}, \ and\ \bibinfo {author} {\bibfnamefont {B.}~\bibnamefont
  {Morosin}},\ }\href {\doibase 10.1107/S0567740873005911} {\bibfield
  {journal} {\bibinfo  {journal} {Acta Crystallographica Section B}\ }\textbf
  {\bibinfo {volume} {29}},\ \bibinfo {pages} {1991} (\bibinfo {year}
  {1973})}\BibitemShut {NoStop}%
\bibitem [{\citenamefont {Togo}\ and\ \citenamefont {Tanaka}(2015)}]{phonopy}%
  \BibitemOpen
  \bibfield  {author} {\bibinfo {author} {\bibfnamefont {A.}~\bibnamefont
  {Togo}}\ and\ \bibinfo {author} {\bibfnamefont {I.}~\bibnamefont {Tanaka}},\
  }\href@noop {} {\bibfield  {journal} {\bibinfo  {journal} {Scr. Mater.}\
  }\textbf {\bibinfo {volume} {108}},\ \bibinfo {pages} {1} (\bibinfo {year}
  {2015})}\BibitemShut {NoStop}%
\bibitem [{\citenamefont {Li}\ \emph {et~al.}(2014)\citenamefont {Li},
  \citenamefont {Carrete}, \citenamefont {Katcho},\ and\ \citenamefont
  {Mingo}}]{ShengBTE}%
  \BibitemOpen
  \bibfield  {author} {\bibinfo {author} {\bibfnamefont {W.}~\bibnamefont
  {Li}}, \bibinfo {author} {\bibfnamefont {J.}~\bibnamefont {Carrete}},
  \bibinfo {author} {\bibfnamefont {N.~A.}\ \bibnamefont {Katcho}}, \ and\
  \bibinfo {author} {\bibfnamefont {N.}~\bibnamefont {Mingo}},\ }\href
  {\doibase http://dx.doi.org/10.1016/j.cpc.2014.02.015} {\bibfield  {journal}
  {\bibinfo  {journal} {Computer Physics Communications}\ }\textbf {\bibinfo
  {volume} {185}},\ \bibinfo {pages} {1747 } (\bibinfo {year}
  {2014})}\BibitemShut {NoStop}%
\bibitem [{\citenamefont {Sadasivam}\ \emph {et~al.}(2016)\citenamefont
  {Sadasivam}, \citenamefont {Ye}, \citenamefont {Charles}, \citenamefont
  {Miao}, \citenamefont {Feser}, \citenamefont {Kubis},\ and\ \citenamefont
  {Fisher}}]{sadasivam2016thermal}%
  \BibitemOpen
  \bibfield  {author} {\bibinfo {author} {\bibfnamefont {S.}~\bibnamefont
  {Sadasivam}}, \bibinfo {author} {\bibfnamefont {N.}~\bibnamefont {Ye}},
  \bibinfo {author} {\bibfnamefont {J.}~\bibnamefont {Charles}}, \bibinfo
  {author} {\bibfnamefont {K.}~\bibnamefont {Miao}}, \bibinfo {author}
  {\bibfnamefont {J.~P.}\ \bibnamefont {Feser}}, \bibinfo {author}
  {\bibfnamefont {T.}~\bibnamefont {Kubis}}, \ and\ \bibinfo {author}
  {\bibfnamefont {T.~S.}\ \bibnamefont {Fisher}},\ }\href@noop {} {\bibfield
  {journal} {\bibinfo  {journal} {arXiv preprint arXiv:1609.03063}\ } (\bibinfo
  {year} {2016})}\BibitemShut {NoStop}%
\bibitem [{\citenamefont {Sadasivam}\ \emph {et~al.}(2015)\citenamefont
  {Sadasivam}, \citenamefont {Waghmare},\ and\ \citenamefont {Fisher}}]{RN22}%
  \BibitemOpen
  \bibfield  {author} {\bibinfo {author} {\bibfnamefont {S.}~\bibnamefont
  {Sadasivam}}, \bibinfo {author} {\bibfnamefont {U.~V.}\ \bibnamefont
  {Waghmare}}, \ and\ \bibinfo {author} {\bibfnamefont {T.~S.}\ \bibnamefont
  {Fisher}},\ }\href {\doibase 10.1063/1.4916729} {\bibfield  {journal}
  {\bibinfo  {journal} {Journal of Applied Physics}\ }\textbf {\bibinfo
  {volume} {117}},\ \bibinfo {pages} {134502} (\bibinfo {year}
  {2015})}\BibitemShut {NoStop}%
\bibitem [{\citenamefont {Kosevich}(1995)}]{RN20}%
  \BibitemOpen
  \bibfield  {author} {\bibinfo {author} {\bibfnamefont {Y.~A.}\ \bibnamefont
  {Kosevich}},\ }\href {\doibase 10.1103/PhysRevB.52.1017} {\bibfield
  {journal} {\bibinfo  {journal} {Physical Review B}\ }\textbf {\bibinfo
  {volume} {52}},\ \bibinfo {pages} {1017} (\bibinfo {year}
  {1995})}\BibitemShut {NoStop}%
\bibitem [{\citenamefont {Hohensee}\ \emph {et~al.}(2015)\citenamefont
  {Hohensee}, \citenamefont {Wilson},\ and\ \citenamefont {Cahill}}]{RN21}%
  \BibitemOpen
  \bibfield  {author} {\bibinfo {author} {\bibfnamefont {G.~T.}\ \bibnamefont
  {Hohensee}}, \bibinfo {author} {\bibfnamefont {R.~B.}\ \bibnamefont
  {Wilson}}, \ and\ \bibinfo {author} {\bibfnamefont {D.~G.}\ \bibnamefont
  {Cahill}},\ }\href {\doibase 10.1038/ncomms7578} {\bibfield  {journal}
  {\bibinfo  {journal} {Nat Commun}\ }\textbf {\bibinfo {volume} {6}},\
  \bibinfo {pages} {6578} (\bibinfo {year} {2015})}\BibitemShut {NoStop}%
\bibitem [{\citenamefont {Asheghi}\ \emph {et~al.}(2002)\citenamefont
  {Asheghi}, \citenamefont {Kurabayashi}, \citenamefont {Kasnavi},\ and\
  \citenamefont {Goodson}}]{RN23}%
  \BibitemOpen
  \bibfield  {author} {\bibinfo {author} {\bibfnamefont {M.}~\bibnamefont
  {Asheghi}}, \bibinfo {author} {\bibfnamefont {K.}~\bibnamefont
  {Kurabayashi}}, \bibinfo {author} {\bibfnamefont {R.}~\bibnamefont
  {Kasnavi}}, \ and\ \bibinfo {author} {\bibfnamefont {K.~E.}\ \bibnamefont
  {Goodson}},\ }\href {\doibase 10.1063/1.1458057} {\bibfield  {journal}
  {\bibinfo  {journal} {Journal of Applied Physics}\ }\textbf {\bibinfo
  {volume} {91}},\ \bibinfo {pages} {5079} (\bibinfo {year}
  {2002})}\BibitemShut {NoStop}%
\end{thebibliography}
%

\end{document}